\documentclass[]{interact}

\usepackage{epstopdf}
\usepackage{subfigure}
\usepackage{graphicx}
\usepackage[square,numbers]{natbib}
\theoremstyle{plain}

\theoremstyle{definition}

\theoremstyle{remark}

\begin{document}


\title{The thermodynamics of clocks.}

\author{
\name{G.~J. Milburn\thanks{CONTACT G.~J. Milburn. Email: milburn@physics.uq.edu.au}}
\affil{Centre for Engineered Quantum Systems, School of Mathematics and Physics,\\ The University of Queensland, Australia 4072.}
}

\maketitle

\begin{abstract}
All clocks, classical or quantum, are open non equilibrium  irreversible systems subject to the constraints of thermodynamics.  Using examples I show that these constraints necessarily limit the performance of clocks and that good clocks require large energy dissipation.  For periodic clocks, operating on a limit cycle, this is a consequence of phase diffusion.  It is also true for  non periodic clocks (for example, radio carbon dating) but due to telegraph noise not to phase diffusion. In this case a key role is played by accurate measurements that decrease entropy, thereby raising the free energy of the clock, and requires access to a low entropy reservoir. In the quantum case, for which thermal noise is replaced by quantum noise (spontaneous emission or  tunnelling), measurement  plays an essential role for both periodic and non periodic clocks. The paper concludes with a discussion of the Tolman relations and  Rovelli's thermal time hypothesis in terms of clock thermodynamics. 
\end{abstract}


\section{Introduction}
A clock is a machine and like all machines subject to the laws of  thermodynamics. In popular discussions, a clock is often presented as the epitome of a reversible and predictable dynamical system. Nothing could be further from the truth. As I will explain, a careful examination of the requirements for a clock indicates that a clock cannot be reversible and in fact requires dissipation for correct operation. I will use a number of examples, both classical and quantum, to illustrate this. I will also discuss irreversible clocks such as radio carbon dating and Mach's temperature clock. A unified description can be given by focusing on how clocks, as non equilibrium steady state systems, exploit sources of low entropy energy (high free energy). 

There is a bewildering range of clocks, from bio-chemical to astronomical, from mechanical to stochastic, and it is difficult to give a general definition of a clock to cover all instances. A standard dictionary definition states ``a clock is a device to measure time", which begs a very much bigger and enduring question. Other definitions emphasise the repetitive or periodic nature of clock states, yet radio carbon dating is a clock that is anything but periodic as was the water clock used by Galileo. 


I will take the view that a clock is a physical device used to coordinate local coincidences of physical events. This captures the central features of time in General Relativity; proper time is local and coordinates relationships between local physical events.  Under this definition clocks can be used to plan ahead by referring to future coincidences, to sequence historical events (radio-carbon dating), or to synchronize events in different locations for navigation. This definition also shares some of the characteristics of distance measurements in enabling us to speak about `here and now'. If clocks do measure time, this definition captures the relational character of time.  

In order to use a physical system to coordinate coincidences, we count the number of times a recurring physical event is observed to occur in the clock between coincident events happening here and now and coincident events here and in the future. We can add to the basic clock a device that keeps the count and displays the number.  Clocks are all about counting. 

Length measurements are also about counting: count the number of times a ruler is placed, end to end,  between one point and another. Good rulers are those systems that do not change their length when used in this way. If they did change, it would cause the count to vary for repeated measurements of the same distance. Likewise, good clocks are those for which the number of counts between periodic coincidences does not fluctuate. Of course in reality there is no perfect ruler; depending on the material used the length of the ruler could fluctuate a little or a lot. Likewise, there is no perfect clock. How good a ruler or a clock needs to be depends on the nature physical events that we seek to coordinate.

\section{The essential irreversible character of periodic clocks.}
I will begin with periodic clocks.  As any rudimentary textbook on clocks will describe\cite{Jes99}, the simplest such devices are driven, periodic systems, perhaps stabilised by feedback and with some counting mechanism to track the number of repetitions of a periodic variable. 

The  pendulum clock with the  escapement mechanism revolutionised the mechanical clock following its invention by Huygen's in 1656. It comprises a pendulum that, for small oscillations in the absence of dissipation, will oscillate at constant amplitude and frequency (the natural frequency) determined by its length and acceleration due to gravity.  But a pendulum once set in motion will eventually stop oscillating due to friction between the support and the rod, collisions with air molecules or other uncontrollable interactions with the environment.  To sustain an oscillation the pendulum must be driven and this is the role of the escapement in early pendulum clocks.  

The purpose of the escapement mechanism is to input energy to the pendulum to counteract the effects of dissipation. With careful design, the driven pendulum  will settle into a steady oscillation at or close to its natural frequency. We refer to this sustained oscillation as a limit-cycle; a stable steady state of sustained oscillatory motion. Limit cycles are common in driven, dissipative non linear systems.  Dissipative attractors, such as a limit cycle, ensure clocks based on this phenomenon are necessarily stochastic dynamical  systems; a necessary consequence of the fluctuation-dissipation theorem\cite{FDtheorem2}. 

\subsection{Dissipation and noise in the escapement pendulum clock.}

\begin{figure}[ht]
\centering
\includegraphics[scale=0.6]{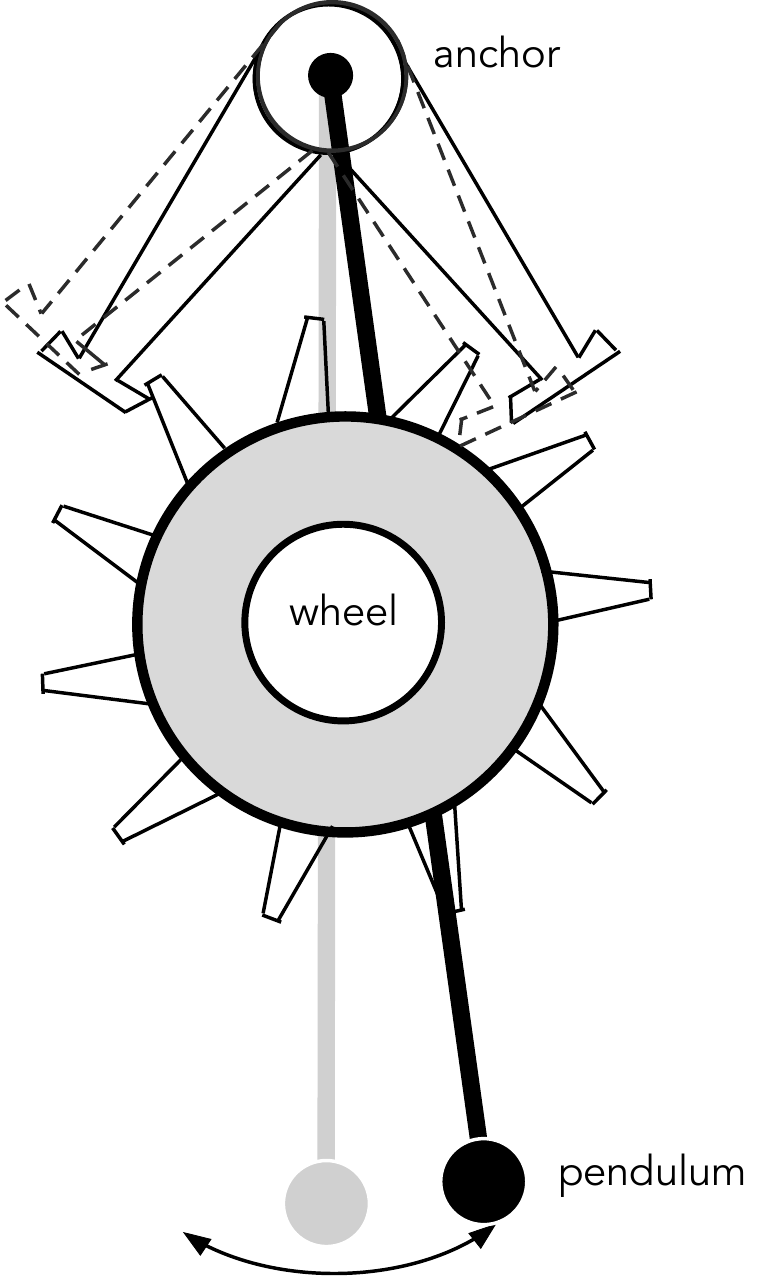}
\caption{A schematic anchor and escape wheel. The falling mass that provides a constant torque to the wheel is not shown. As the pendulum unlocks the anchor from its right hand or left hand position, it imparts a small impulsive torque to the pendulum. The sequence is: left-lock $\rightarrow$ clockwise kick $\rightarrow$ free $\rightarrow $ right-lock $\rightarrow$ anticlockwise kick $\rightarrow$ free $\rightarrow $ left-lock. Initial transients are not shown.  }
\label{graham-escapement}
\end{figure}

A pendulum, for small oscillations, is a linear dynamical system so how can a limit cycle occur? It is the driving mechanism of the escapement itself that is responsible for the non linearity. The magnitude of the impulse delivered depends non linearly on the angular displacement and possibly the angular velocity. Non linearity is central to clock design.  I will use the Graham escapement mechanism as an example\cite{DuXie,animation} schematically indicated in Fig.(\ref{graham-escapement}) based on the ``wheel and anchor" mechanism. 

 I will adapt the treatment of \cite{Hoy14} to a Hamiltonian phase-space description. For a pendulum the phase space variables are angular displacement, $\theta$, and the angular momentum, $p$. When dissipation and fluctuations are included  Hamilton's equations are modified to become Langevin equations. In this case,
\begin{eqnarray}
\dot{\theta} & = & \frac{p}{m\omega l}\\
\dot{p} & = & -mg \theta-\gamma p+K(\theta,p)+\xi(t)
\end{eqnarray}
where the $\cdot$ indicates a time derivative,  $m$ is the mass, $\omega=\sqrt{g/l}$ is the natural frequency of the pendulum of length $l$ and  $\gamma$ is the frictional damping rate. The fluctuating force term $\xi(t)$ is a white noise process of mean zero and (stationary) two-time correlation function
$\overline{\xi(t)\xi(t+\tau)}= D\delta(\tau)$. The fluctuation-dissipation theorem requires $D=2m\gamma k_BT$. The term $K(\theta,p)$ describes small impulsive kicks delivered by the design of the escapement\cite{Hoy14}. 

\begin{figure}[ht]
\centering
\includegraphics[scale=0.25]{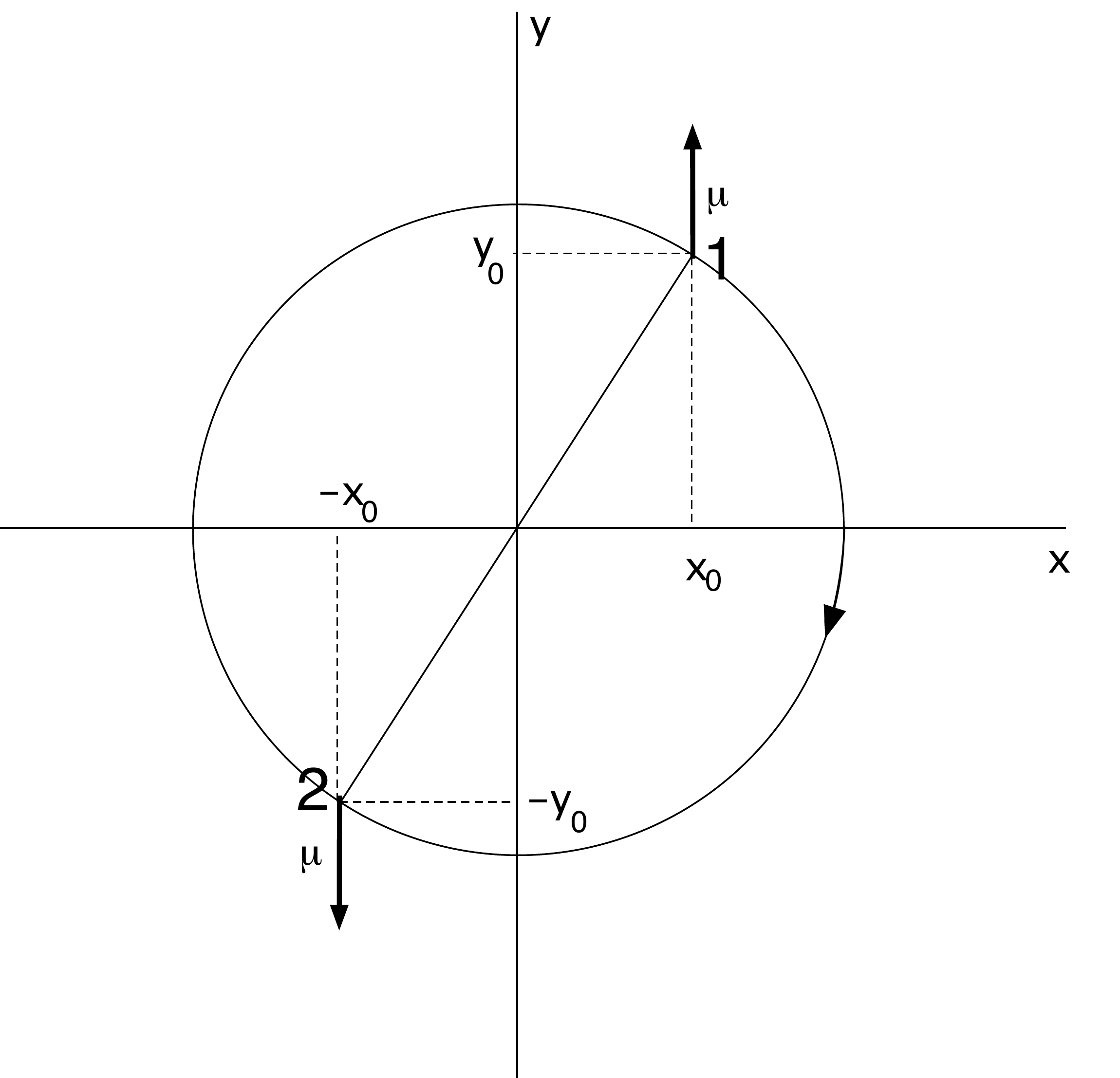}
\caption{A phase-space representation of momentum kicks provided to the pendulum every half cycle by the escapement mechanism with $\tan\psi_0=y_0/x_0$. At point labeled 1, the impulse switches  $-\mu\rightarrow +\mu$ while at point labeled 2 the impulse switches $+\mu\rightarrow -\mu$ }\label{fig1}
\end{figure}

\begin{figure}[ht]
\centering
\includegraphics[scale=0.5]{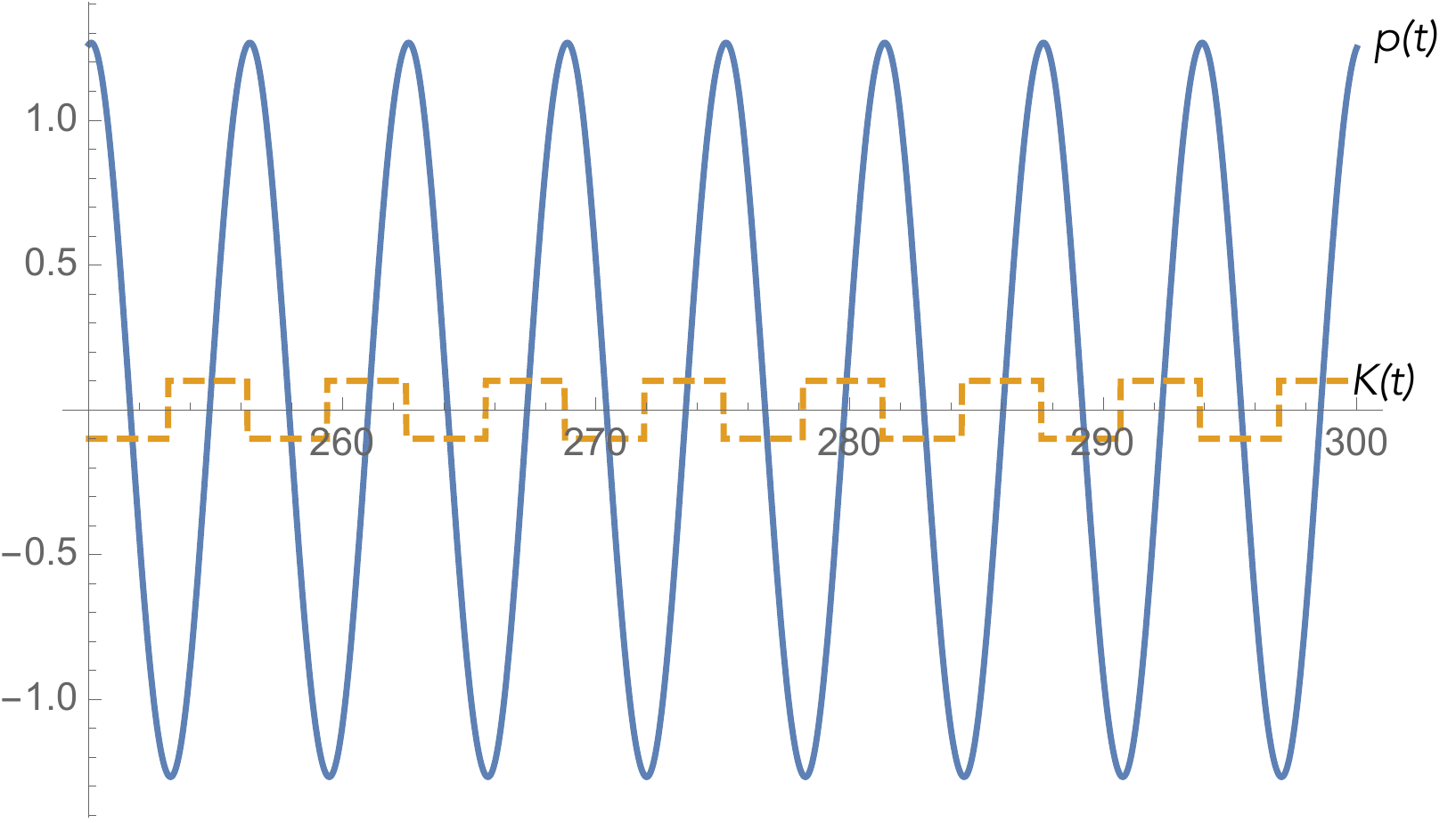}
\caption{The  angular displacement (solid) and the impulse function (dashed) as a function of time for the kicked pendulum with $\sin\psi_0=0.1\ \Gamma=0.1\ \mu=0.1$.  }\label{fig2}
\end{figure}

It is simpler to move to dimensionless variables defined by $x=\theta$ and $y=p/(m\omega l)$ and a dimensionless time $\tau=\omega t$, so that the equations take the form
\begin{eqnarray}
\label{kicked-model-langevin}
\dot{x} & = & y \\
\dot{y} & = & -x-\Gamma y+K(x,y)+\eta(t)
\label{kicked-model}
\end{eqnarray}
where $\Gamma=\frac{\gamma}{m\omega^2l}$ and $\eta(t) = \frac{\xi(t)}{m\omega^2 l}$. 
For the kicks we take
\begin{equation}
K(x,y)=-\mu\ {\rm sign}(\sin\psi_0\ x-\cos\psi_0\ y )
\end{equation}
where $\psi_0$ is fixed by the design of the escapement and $\mu$ has units of frequency.  This is a non linear function of the phase space variables. Note that as we are using the quadratic approximation for the pendulum potential function we need to ensure that $x_0 << \pi/2$. We can easily remove this approximation, see \cite{Hoy14}.  A phase-space picture of this impulse is shown in Fig. \ref{fig1}. It provides  constant kick to the momentum once every half period.  
In Fig,\ref{fig2} we plot the dimensionless momentum, $y$, as a function of time together with the impulse function $K(x,y)$, in units where $\mu=1$.  The stable limit cycle is apparent.

 In order to extract the limit cycle, define new variables, $r,\psi$, by $x=r\cos\psi$ and $y= r\sin\psi$. 
The time scale of the decay onto the limit cycle (the $r(t)$ dynamics) is slower than the oscillation frequency on the limit cycle (the $\psi(t)$ dynamics). This implies that we can average over the oscillatory faster time scale which is equivalent to averaging over $\psi$ from $-\pi-\psi_0$ to $\pi-\psi_0$,  the equation for the radial coordinate after averaging is
\begin{equation}
\label{radial-motion}
\dot{r} = -\frac{\Gamma r}{2}+\frac{2\mu}{\pi}\cos\psi_0
\end{equation}
It is now clear that there is a stable fixed point in the radial coordinate given by, 
\begin{equation}
\label{radial-fp}
r_*=\frac{4\mu}{\pi\Gamma}\cos\psi_0
\end{equation}
The dynamics for the phase is 
\begin{equation}
\dot{\psi} = -1+\frac{2\mu}{r\pi}\sin\psi_0
\end{equation}
The frequency is shifted from that of the free oscillator and only approaches the free oscillator in a limit such that the size of the impulsive kicks goes to zero. 

In scaled units we define the energy of the oscillator as 
\begin{equation}
    E=\frac{1}{2}(x^2+y^2)=\frac{1}{2}r^2
\end{equation}
If we time-average this over the fast oscillations on the limit cycle we find that
\begin{equation}
    \dot{E}=-\Gamma E+ \frac{2\sqrt{2E}\mu}{\pi}\cos\psi_0
\end{equation}
The last term is the average rate at which work is done by the escapement on the pendulum. 
The limit cycle forms when the energy lost due to friction is balanced by the work done by the escapement. 
The important feature is that the work done increases with the size of the limit cycle. In the steady state, the average energy on the limit cycle is
\begin{equation} 
\overline{E}=\frac{8\mu^2}{\Gamma^2\pi^2}\cos^2\psi_0
\end{equation}
Of course this result depends on the assumption that the dynamics is deterministic; noise is neglected.

A dynamical system subject to dissipation must necessarily experience noise. This is a direct consequence of the fluctuation-dissipation theorem\cite{FDtheorem,FDtheorem2}.  The inclusion of white noise in non linear dynamical problems of this kind is not easy. The nonlinear transformation to radial and phase variables leads to non linear noise in both $\dot{r},\dot{\psi}$.  I will simply quote the result in \cite{Zhu87} for Ito stochastic differential equations after phase averaging,
\begin{eqnarray}
dr & = & [-\frac{\Gamma r}{2}+\frac{2\mu}{\pi}\cos\psi_0+\frac{D'}{4r}]dt+\sqrt{D'/2}\ dW_1(t) \\
d\psi & = & [-1+\frac{2\mu}{r\pi}\sin\psi_0]dt+ \frac{\sqrt{D'/2}}{r}\ dW_2(t)
\end{eqnarray}
where $dW_1(t), dW_2(t)$ are independent Wiener increments (random variables with zero mean and variance $dt$) and the diffusion constant $D'$ is given in terms of the dimensional diffusion constant $D$ as 
\begin{equation}
D'=\frac{D\omega}{m^2g^2}=\frac{2\omega \gamma k_B T}{mg^2}
\end{equation}
Further details on stochastic differential equations, Wiener increments and transforming from Langevin equations to Ito equations may be found in \cite{Gar83}.

Two important effects are apparent:  a small shift in the the radial fixed point and a non linear diffusion in the phase. The radial fixed point is changed to
\begin{equation}
r_*= \frac{2\mu}{\pi\Gamma}\cos\psi_0+\sqrt{\frac{4\mu^2}{\pi^2\Gamma^2}\cos^2\psi_0+D'/4}
\end{equation}
We can linearise the effect of phase noise by setting setting $r\rightarrow r_*$ in the stochastic term for $d\psi$. This leads to the significant fact that the phase diffusion gets slower the larger the limit cycle. For typical pendulum clocks run at room temperature the phase diffusion rate is very small indeed and it is no surprise that this fundamental source of noise is not at all significant for such clocks. 

However there are cases, especially for quantum clocks, where the phase noise on the limit cycle does limit the performance of the clock. The analysis begins by linearising the noise when the clock has settled onto the limit cycle.

\subsection{Noise and period fluctuations}
\label{period-fluctuations}
To understand the effect of noise it is useful to write the kick function directly in terms of the phase variable on the limit cycle as
\begin{equation}
    K(t)=-\mu\ {\rm sign}[\cos(\psi(t)+\psi_0)]
\end{equation}
In the deterministic case, $\psi(t)=\tilde{\omega}t$ and this function is periodic on $-\pi/2-\psi_0$ to $3\pi/2-\psi_0$. The transitions between $\pm\mu$ occur at regular intervals separated in time by  $\pi/\tilde{\omega}$. As the noise adds a random Gaussian variable to $\psi(t)$ over time $t$, these transitions become irregular; some a little early, some a little late. This suggests that the function $K(t)$ is developing the characteristics of a random telegraph signal (RTS). However it cannot be a simple RTS with constant transition rates as the transitions are most likely to occur at particular moments in time near the end points of the oscillation. \begin{figure}[ht]
\centering
\includegraphics[scale=0.4]{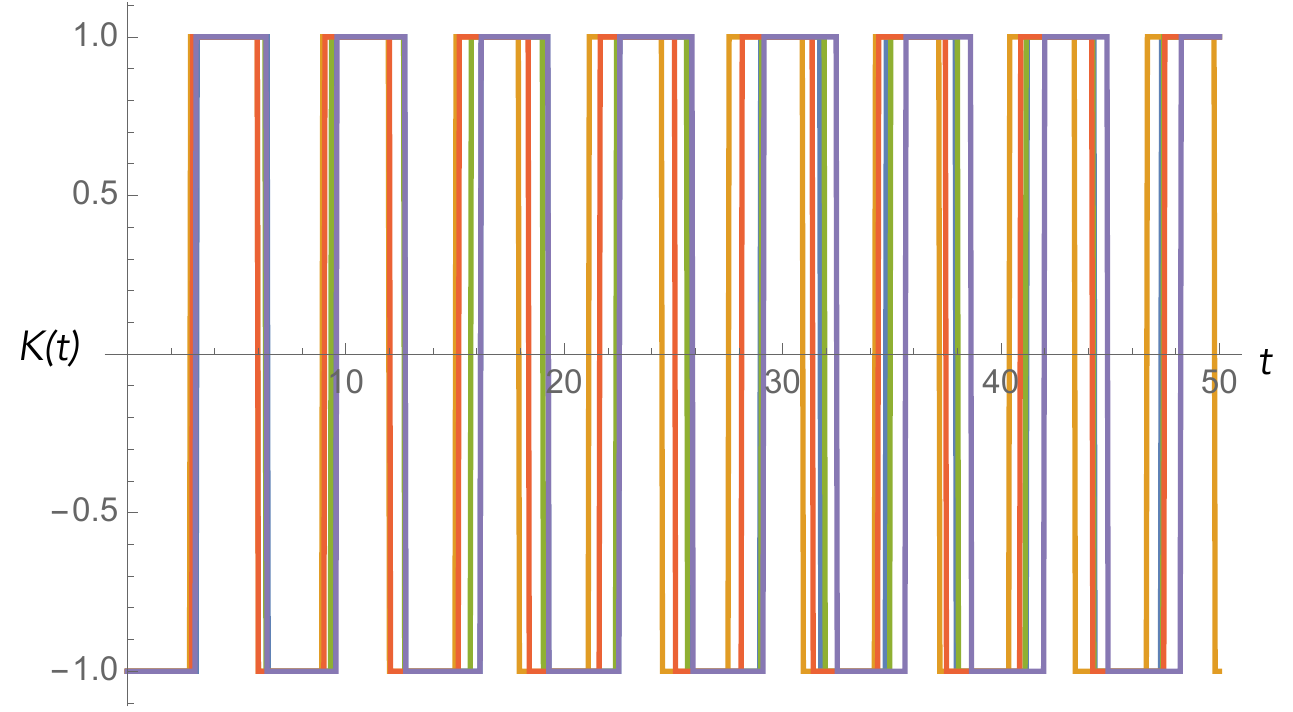}
\caption{A plot of five sample trajectories of $K(t)$ with phase diffusion as a function of $t$ with $\sin(\psi_0)=0.1$,$\mu=\Gamma=1.0$ and $\sigma=0.1$   }\label{random-pase-sim}
\end{figure}
In Fig.(\ref{random-pase-sim}) I show five sample trajectories obtained by solving the Ito stochastic differential equation for the phase on the limit cycle
\begin{equation}
    d\psi(t)=-\tilde{\omega}dt+\frac{\sigma}{r_*} dW(t)
\end{equation}
and evaluating $K(t)$. The fluctuations in the switching times are clear.

\begin{figure}[ht]
\centering
\includegraphics[scale=0.4]{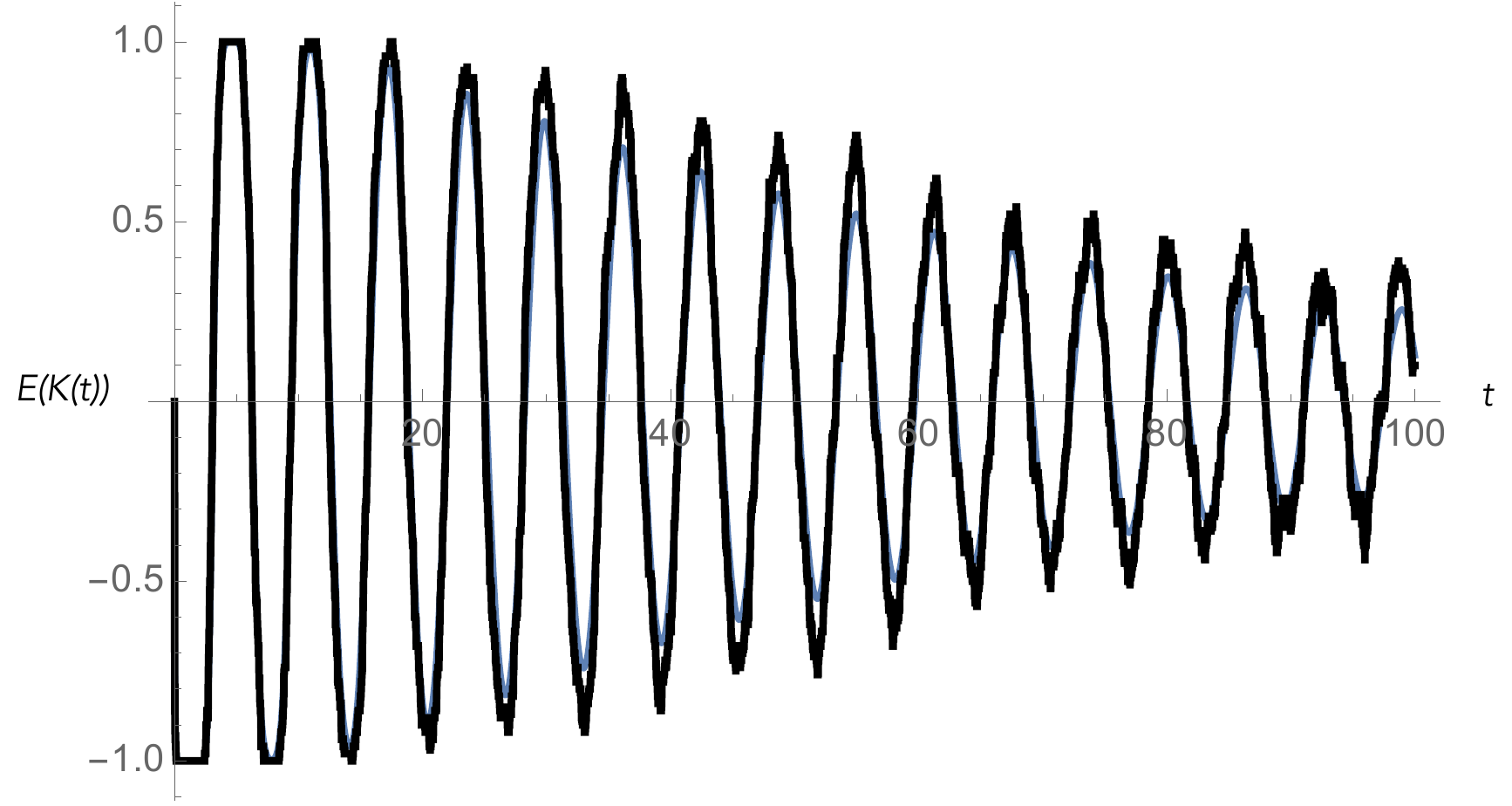}
\caption{The ensemble average over 100 samples of the process $K(t)$ as a function of $t$ with $\psi_0=0$, $\tilde{\omega}=1$,$\mu=\Gamma=1.0$ and $\sigma=0.2$. Also shown is the analytic expression given  in Eq. (\ref{average-kick})}\label{kick-ensemble-av}
\end{figure}

Expanding the sign function in a Fourier series, the ensemble average of the $K(t)$, with $\psi_0=0$, on the limit cycle can be shown to be given by 
\begin{equation}
\label{average-kick}
    E(K(t))=\frac{4\mu}{\pi}\sum_{k=1}^\infty e^{-\sigma^2 t(2k-1)^2/(2r_*^2)}\frac{\sin((2k-1)\tilde{\omega} t)}{2k-1}
\end{equation}
In Fig.(\ref{kick-ensemble-av}) I plot the ensemble average computed from 100 trials of the process shown in Fig.(\ref{random-pase-sim}) as well as the analytical expression given in Eq.(\ref{average-kick}). 

To estimate how the noise impacts the time keeping, fix the interval of external time that passes as $T$ and ask how many `ticks' of the clock occur in this time. In the clock literature this approach is known as time domain characterisation, as opposed to the frequency domain where the emphasis is on noise power spectral density\cite{Rutman}. The escapement force can be used to monitor this count. Each cycle of the square wave kick function corresponds to one tick of the clock.  If there were  no noise, a count of $M$ cycles  corresponds to a  period in conventional time units of  $T/M$. But in the presence of noise the count $M$ and thus the assigned period fluctuates from one cycle to the next.
\begin{figure}[ht]
\centering
\includegraphics[scale=0.4]{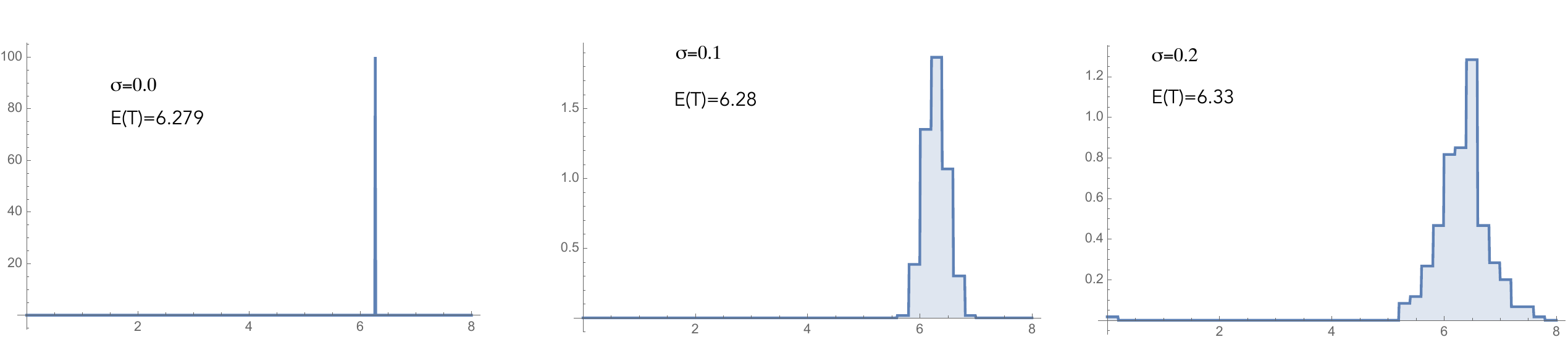}
\caption{A histogram distribution of the same cycle of $K(t)$ over 300 trials   with $A=0.0$, $\gamma=0.1,\mu=0.1$, $\omega=1.0$.Values for the mean period is $E(T)$}
\label{period-distrib}
\end{figure}
 Alternatively one can sample the time for one cycle, that is to say the period $T$, over many trials. In Fig. (\ref{period-distrib} )  I plot a histogram distribution of this period  in $500$ trials for three different value of the noise level.  The zero noise case gives a period of $6.35$ which is the correct value for these parameters. 
 
 In this linearised noise treatment it remains the case that the energy on the limit cycle does not change. However the work done by the escapement and hence the energy lost to friction fluctuates from one cycle of the escapement to the next. To see this observe that it follows from the equations of motion that the work, $W$, done by the escapement occurs at the rate  $K(x,y)y$. Let one cycle of the clock start with a switch from $K=-\mu$ to $K=+\mu$. The the work done by the escapement over one cycle is 
 \begin{equation}
     W=\mu r_*\left [\int_{\tau_+}\sin(\psi(t))dt +\int_{\tau_-}|\sin(\psi(t))|dt\right ]
 \end{equation}
where $\tau_\pm$ is the time spent in the state $K=\pm\mu$. As these intervals are random variables the work done in each cycle is a random variable. The heat dissipated on each is given by cycle $Q=W/\Gamma$, so fluctuations in $W$ correspond to fluctuations in the heat dissipated. Putting it together,  there is a relationship between fluctuations in the clock period and the fluctuations in the heat dissipated by the clock. 

In the deterministic case the work done averaged over each cycle is related to the period by 
\begin{equation}
\label{work-v-period}
    W=\left (\frac{2\mu r^*\cos\psi_0}{\pi} \right )T
\end{equation}
A simple estimate of the average heat dissipated per cycle   is 
\begin{equation}
 Q=\left (\frac{2 \mu r^*\cos\psi_0}{\pi\Gamma} \right )T
\end{equation}
Including noise we expect both $Q$ and $T$ to fluctuate from one cycle to the next. 

\begin{figure}[ht]
\centering
\includegraphics[scale=0.4]{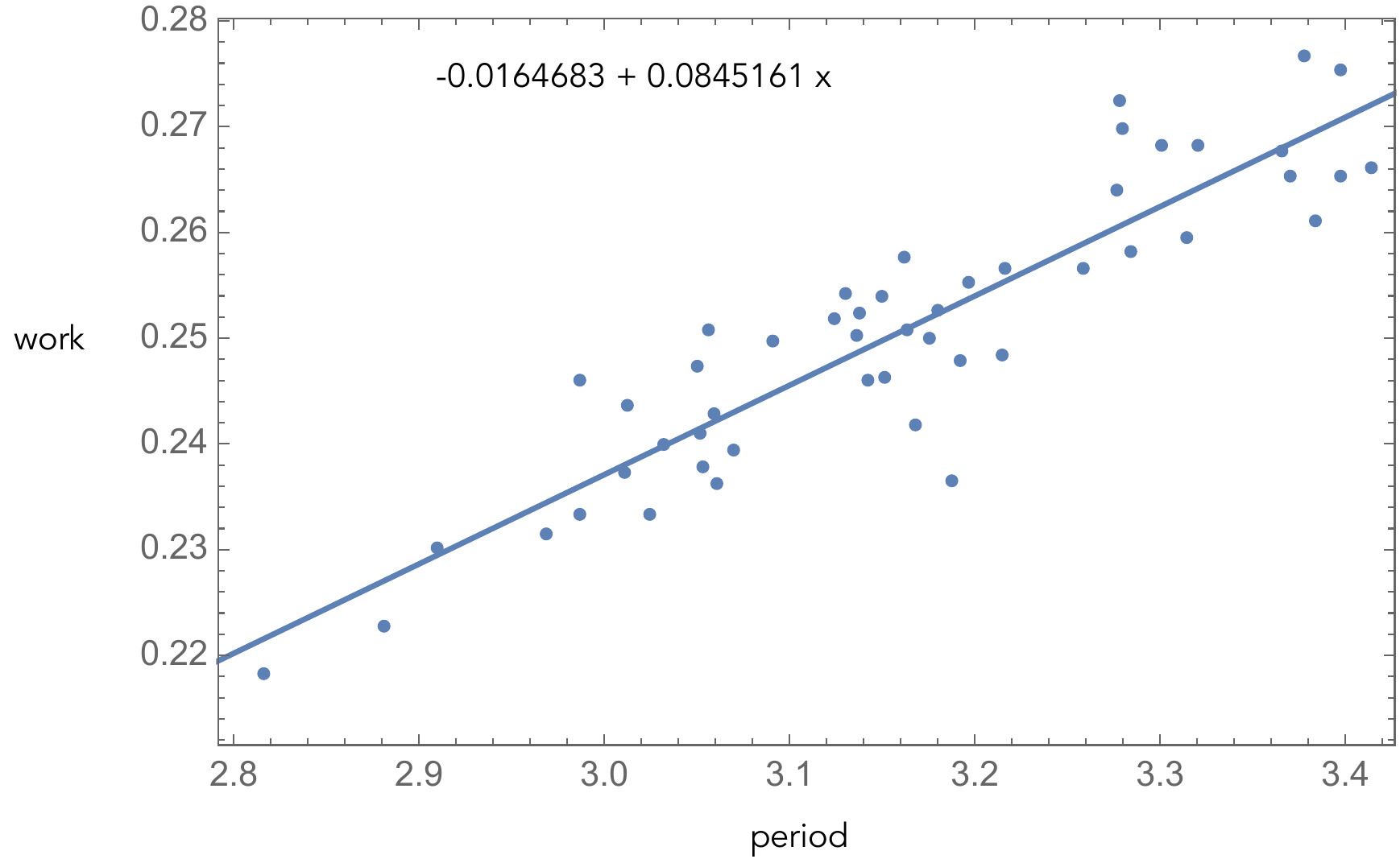}
\caption{A plot of the work done versus period for the first half cycle in 50 sample trajectories with $\sin\psi_0=0.2, \gamma=0.1,\mu=0.1, \omega=1.0,\sigma =0.1$. The linear fit is  $-0.0164683 + 0.0845161 x$ }
\label{work-period}
\end{figure}

In Fig.(\ref{work-period}) I plot the result of 50 simulations for the work done on the first half cycle versus the corresponding period.  The work done is computed numerically from the integral 
\begin{equation}
    W_k=\mu r^*\int_0^{T_k/2} dt \sin\psi(t))
\end{equation}
by numerically solving the Ito stochastic differential equations for $d\psi(t)$ using the Ito solver in Mathematica. A linear fit indicates a slope of $0.085$ while Eq. (\ref{work-v-period}) for these parameters gives a slope of $0.078$.

The fluctuations in $Q$ thus determine the fluctuations in the period and thus the quality of the clock. Noting  that the fluctuations in the period depend on the phase diffusion noise and this scales inversely with the size of the limit cycle $r_*$, while the average heat lost per cycle scales with the size of the limit cycle, the more accurate the clock the greater is the heat dissipated. This important thermodynamic constraint on the accuracy of clocks was first presented by  Erker at al. \cite{Erker2017} in the context of an abstract quantum model. We see here it is a consequence of the dissipative nature of a simple classical periodic clock.  It also follows that on the limit cycle $Q/T=4\mu r_*$ , a constant on all cycles. This implies that the fractional error in heat dissipated and clock frequency are related by 
\begin{equation}
  \frac{\delta Q}{Q}  =\frac{\delta \omega}{\omega}
\end{equation}
A similar result occurs in the context of non periodic thermal clocks as I discuss in section (\ref{non-periodic-clocks}). In section (\ref{quantum-periodic-clocks}) I will consider quantum periodic clocks. There the noise on the limit cycle arises from a fundamental quantum sources of noise: spontaneous emission and quantum tunnelling. These processes do not go to zero at zero temperature and thus quantum thermodynamics limits the accuracy of all clocks.

\subsection{Dissipation and noise in the quartz oscillator clock.}
As a second example of a well known periodic clock I will consider the electronic quartz oscillator clock.   A shortcoming of the treatment of the escapement pendulum clock in the previous section is the omission of the back-action of the pendulum on the escapement: the kick function was simply imposed. The quartz oscillator clock gives an opportunity to move beyond this. The two key discoveries that led to the quartz clock are; the discovery of relaxation oscillations (another name for limit cycles) in vacuum tube circuits and the high quality piezo-electric oscillations exhibited by small crystals of quartz. A good summary of the history is given by the inventor, Warren Marrison \cite{Mar48}. 

\begin{figure}[ht]
\centering
\includegraphics[scale=0.6]{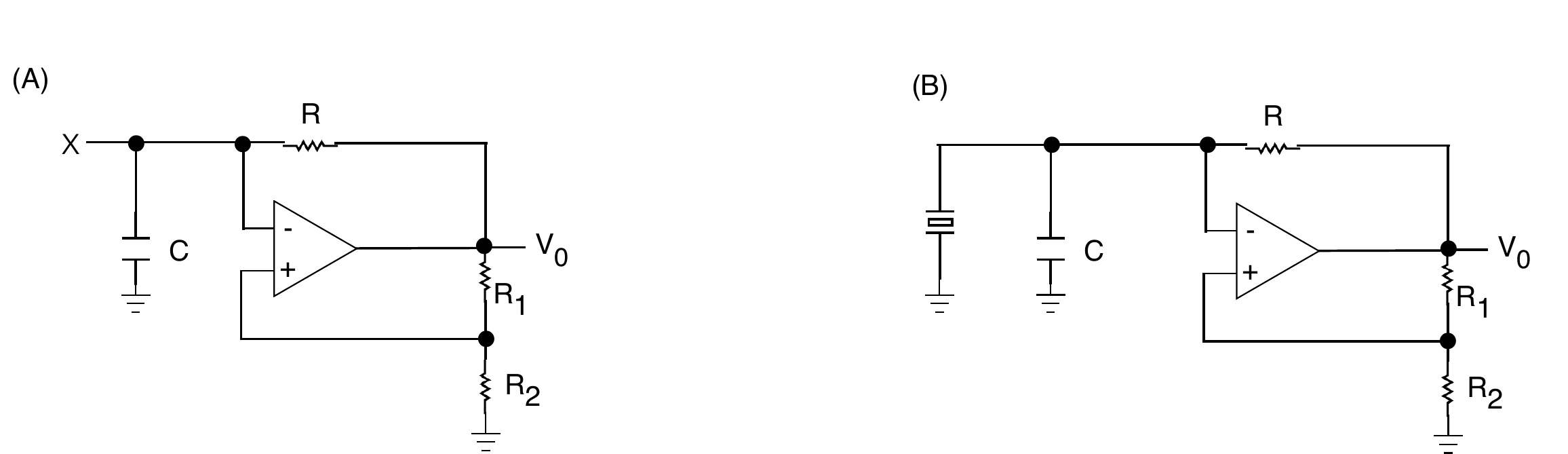}
\caption{(A). A Schmidtt trigger Op Amp. with feedback configured as an astable oscillator (B) A quartz piezo electric oscillator coupled to an astable oscialltor}\label{fig3}
\end{figure}

I will focus here on the relaxation oscillations exhibited in a feedback circuit to form a Schmidtt trigger. A typical Schmidtt trigger has the circuit diagram shown in Fig. \ref{fig3}(A). 
The Op Amp is configured as a comparator.  The input voltage is applied at the inverting input. The non inverting input voltage is determined by a voltage divider comprised of resistors $R_1,R_2$. The initial output of the Op Amp, $V_0$, is at the saturation voltage $V_s$. The voltage at the $+$ terminal is $\beta V_s$ where $\beta =R_1/(R_1+R_2)$. If the input  voltage at the $-$ terminal ramps up from a low value and exceeds $\beta V_s$,  the  output switches to $-V_s$. Now ramping  the voltage at the $-$ terminal back down, the output will switch back to $+V_s$. 

A phenomenological model for the Schmidtt trigger is given by the equation
\begin{equation}
    \dot{V} = \gamma(1-g(V))e^{-\eta X}-\gamma (1+g(V)) e^{\eta X}
\end{equation}
where $V$ is the output voltage and $X$ is the applied voltage at the inverting terminal.  The exponential dependence on voltage, $X$, reflects an avalanche-like switching between reverse and forward biased transistors inside the device.  At steady state for fixed $X$, the system rapidly decays onto the  steady state output voltage $V_{ss}$ as a function of the input $X$ given implicitly by
\begin{equation}
\label{steady-state-voltage}
   g(V_{ss}) =-\tanh(\eta X)
\end{equation}
Clearly $|g(V_{ss})| \leq 1$. In the case of a Schmidtt trigger built from an OP Amp with feed back as in Fig. (5A)
\begin{equation}
\label{gain-curve}
 g(V)=\frac{V+\tanh\beta V}{1+V\tanh\beta V}  
\end{equation}
where $\beta =AR_2/(R_1+R_2)$, and $A$ is a parameter that determines the quality of the Op. Amp. Typically $A>>1$.  This is equivalent to the model discussed in reference \cite{MacWies}.

\begin{figure}[ht]
\centering
\includegraphics[scale=0.4]{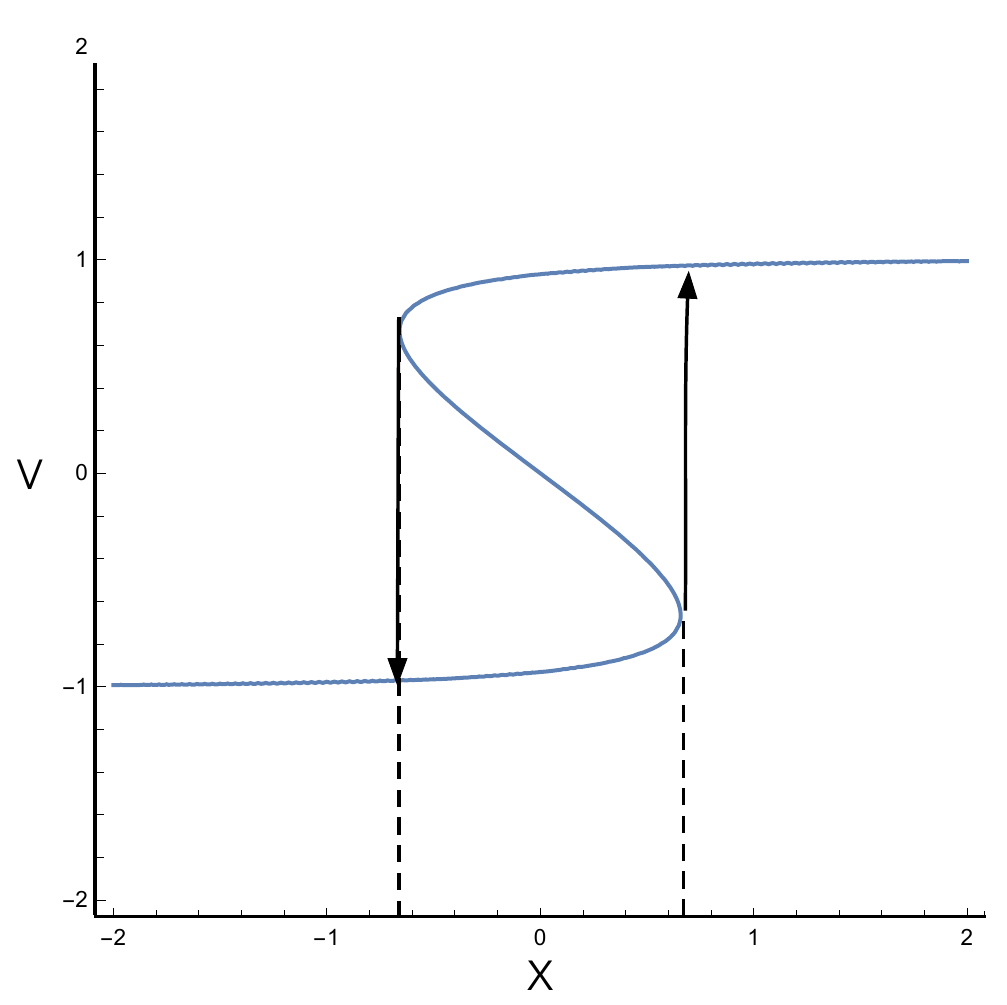}
\caption{Two steady state response functions for a Schmidtt trigger output voltage as a function of the input voltage $X$. $\beta=1.8, \eta=0.6$.}\label{fig4}
\end{figure}
Substituting Eq.(\ref{steady-state-voltage}) into Eq.(\ref{gain-curve}) we can find  $X$ as a function of  $V$,
\begin{equation}
 X=-\frac{\beta}{\eta} V+\frac{1}{2 \eta}\ln\left (\frac{1+V}{1-V}\right )
\end{equation}
The turning points of $X(V)$ are given by $V_c=\pm\sqrt{\frac{\beta-1}{\beta}}$.
The steady state response in shown in Fig.(\ref{fig4}) for the case $\beta=1.8, \eta=0.6$. The required hysteretic response of a Schmidtt trigger is evident.


When we include the quartz oscillator as shown in Fig. \ref{fig3}(B), the input voltage to the Schmidtt trigger is the  voltage on the equivalent capacitor of the quartz oscillator. The equations of motion become,
\begin{eqnarray}
\label{quartz-lc}
\dot{V} & = & \gamma(1-g(V))e^{-\eta X}-\gamma (1+g(V)) e^{\eta X}\\
\dot{X} & = & \omega Y\\
\dot{Y} & = & -\omega X -\kappa Y+\chi V
\end{eqnarray}
   where $V$ is the output voltage of the Schmidtt trigger, $X$ is the voltage on the equivalent circuit capacitor and $Y$ is the flux in the equivalent circuit inductor for the quartz crystal and $\omega$ is the natural frequency of the equivalent LC circuit. The saturation voltage is set to unity. This is not a Hamiltonian system as there are only three equations. This is justified by the large dissipative time constant for the Schmidtt trigger. 
   
\begin{figure}[ht]
\centering
\includegraphics[scale=1.0]{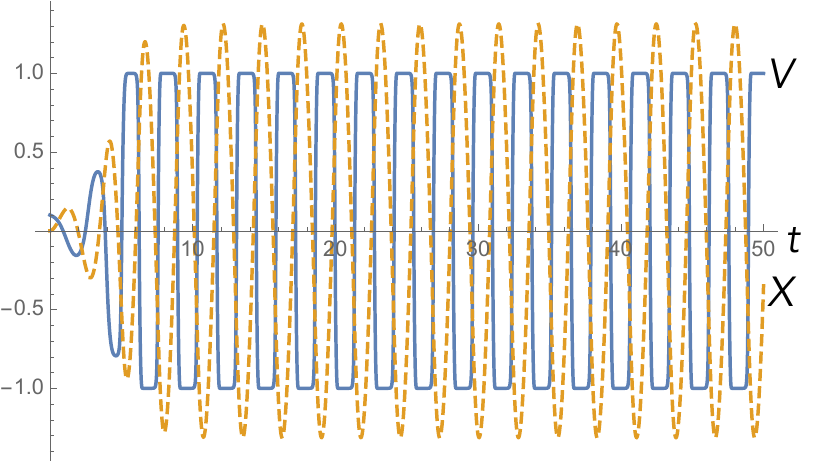}
\caption{The limit cycle trajectories of the Schmidtt trigger output voltage (solid) and the voltage on the quartz crystal (dashed). The parameters are $\omega=1, \gamma= 0.1$, $\eta=8.0$, $\kappa=1.0$, $\beta=1.0$ and $\chi=5.0$.} \label{fig5}
\end{figure}

   A typical limit cycle trajectory for $V$ and $X$ are shown in Fig, \ref{fig5}. The limit cycle occurs via a Hopf bifurcation for values of $\chi$ above a critical value where the fixed point at the origin becomes unstable. 
    Comparing this to the treatment of the escapement clock in the previous section, it is easy to see how this system may be configured as a clock by a simple mechanism to count the number of transitions of the output voltage of the Op Amp.    

On the limit cycle, $g(V(t))=-\tanh(\eta x(t))$ from which it follows that $V(t)=-\tanh(\beta V(t)+\eta x(t))$. Substituting into the equation of motion for $Y$,
\begin{eqnarray}
    \dot{Y} & = & -\omega X-\kappa Y-\chi \tanh(\beta V+\eta x)\\
      & \approx&  -\omega X-\kappa Y-\chi {\rm sign}(x)
\end{eqnarray}
 where the approximation becomes better in the limit that $\eta >> \beta>1$ (noting that $|V(t)|\leq 1$). So this model reduces to the kick-function model discussed in the previous section. 

Noise enters this description in the form of Johnson-Nyquist noise in Ohmic resistance.  A phenomenlogical model of a Schmidtt trigger subject to thermal noise is presented in  \cite{MacWies}. In the model used here this corresponds to adding a fluctuating white noise term to the equation of motion for $X$,
\begin{equation}
\dot{Y} = -\omega X-\kappa Y +\chi V+ \sqrt{D} dW(t)
\end{equation}
where $D$ is a thermal diffusion constant and $dW(t)$ is the Wiener process. The analysis proceeds much as in the previous example. After transients have died out the limit cycle can be described by using the variables $(r,\psi)$.  Linearising the noise on the limit cycle gives a constant phase diffusion. The implications for the accuracy of the clock are unchanged: the more energy is dissipated, the better the clock.

\section{Quantum Periodic clocks.}
\label{quantum-periodic-clocks}

\subsection{Lasers.}
In atomic clocks the primary oscillator is an electric dipole non linear oscillator between a ground and excited state of an atom, although an ensemble of atoms is typically used.  If the energy difference between the ground and excited state is $\epsilon$, the resonance frequency of the atomic oscillator is $\omega=\epsilon/\hbar$.  The driving system, playing the role of the escapement in a mechanical clock, is a laser.  The electric field of the laser drives the atomic dipole transition.  The coupling between the laser and the atomic oscillator can be made in various ways including a direct coupling or a feedback coupling; a distinction that Levine\cite{Levine} refers to as active versus passive respectively.  In atomic clocks both the drive oscillator and the atomic oscillator are nonlinear. 

I will begin with the laser which, as Wiseman has cogently argued, is already a clock\cite{Wiseman-laser-clock}, albeit at a very high frequency. It converts a continuous source of input energy, the `pump', into a coherent oscillation at a (almost) fixed frequency provided the pump power is above a critical threshold. There is an irreducible level of quantum noise in the laser oscillator above threshold due to spontaneous emission. 

The quantum theory of the laser is well established. Although there are a very large variety of lasers I will adopt one model for definiteness: an optical cavity containing a large number of atoms with an atomic transition close to the cavity frequency. A source of incoherent radiation, the `pump',  maintains a population inversion  between the ground and excited state of the transition (i.e. on average, more atoms are in the excited state than the ground state).  

In a competition between the incoherent pumping and loss due to spontaneous and stimulated emission from the laser transition and the optical cavity, the electric field inside the cavity can reach a non equilibrium steady state. This steady state can undergo a non equilibrium phase transition as the pumping is increased above a threshold. Below the threshold the field is Gaussian distributed with average field amplitude of zero. Above the threshold, the field is non-Gaussian with a well defined intensity (Poisson distributed) and slowly diffusing phase with respect to the cavity frequency.  In a semi-classical approach this transition is described as fixed point to limit cycle bifurcation with added noise. 

I will base my discussion on the case in which the cavity loss dominates over atomic decay. In this case the gain medium can be adiabatically eliminated. A good discussion can be found in \cite{Wiseman99} and I will follow that approach. 

The quantum state of the intracavity field can be written in the photon  number basis $|n\rangle$. We define annihilation and creation  operators $a,a^\dagger$ by $a|n\rangle =\sqrt{n}|n-1\rangle$, $a^\dagger|n\rangle =\sqrt{n+1}|n+1\rangle$. It then follows that $a^\dagger a|n\rangle =n|n\rangle$. We call $a^\dagger a $ the photon  number operator.

In the laboratory frame the average cavity field  oscillates at the cavity frequency $\omega_c$. The first step is to move to the interaction picture at the cavity frequency which removes the average field oscillation altogether. 
In the interaction picture the quantum state $\rho$ of the intracavity laser field at any time then satisfies the master equation,
\begin{equation}
\label{laser-me}
  \dot{\rho}  =Gn_s{\cal D}[a^\dagger]\left ({\cal A}[a^\dagger]+n_s\right )^{-1}\rho +\kappa{\cal D}[a]\rho
\end{equation}
where the super-operators are defined by 
\begin{eqnarray}
    {\cal D}[B]\rho  & = & B\rho B^\dagger -\frac{1}{2}(A^\dagger A\rho+\rho A^\dagger A)\\
    {\cal A}[B]\rho & = & \frac{1}{2}(B^\dagger B \rho+\rho B^\dagger B)
\end{eqnarray}
and  $G$ is the small signal gain, $n_s$ is the saturation photon  number and $\kappa$ is the cavity decay rate. The inverse superoperator can be represented as
\begin{equation}
    \left ({\cal A}[a^\dagger]+n_s\right )^{-1}=\int_0^\infty e^{-\beta (a^\dagger a+n_s)/2}\rho e^{-\beta (a^\dagger a+n_s)/2}
\end{equation}
which describes the incoherent pumping process.

I have also assumed operation at optical frequencies to ignore thermal photons entering the cavity. Any noise necessarily arises from quantum fluctuations not thermal fluctuations.  The first term in Eq.(\ref{laser-me}) represents gain, that is to say, the injection of energy into the cavity field, while the second term represents photon loss through the cavity mirrors. 

The probability, $p_n(t)$, to find $n$ photons in the cavity at time $t$ is given by $p_n(t)=\langle n|\rho(t)|n\rangle$. It then follows from Eq. (\ref{laser-me}) that 
\begin{equation}
    \frac{d p_n(t)}{dt}=Gn_s\left [\frac{n}{n+n_s}p_{n-1}-\frac {n+1}{n+1+n_s}p_n\right ]+\kappa(n+1)p_{n+1}-\kappa np_n
\end{equation}
The steady state solution is given by detailed balance\cite{Gar83} as
\begin{equation}
    p_n^{ss}= {\cal N}\frac{(Gn_s/\kappa)^{n+n_s}}{(n+n_s)!}
\end{equation}
where ${\cal N}$ is a normalisation constant. Below threshold $G<\kappa$, the steady state distribution may be approximated by 
\begin{equation}
     p_n^{ss}\approx \frac{1}{1+\bar{n}}\left (\frac{\bar{n}}{\bar{n}+1}\right )^n\ \ \ \ \ \ \  G<\kappa
\end{equation}
which is analogous to a thermal distribution with 
\begin{equation}
    \bar{n}=\frac{G}{\kappa-G}
\end{equation}
Well above threshold $G>>\kappa$ the steady state distribution is Poisson with mean 
\begin{equation}
\label{mean-above}
    \bar{n}=Gn_s/\kappa
\end{equation}

This discussion does not make it particularly transparent why we should regard a laser as a clock.  Indeed the steady state has no dynamics at all. In order to resolve this we need to study the dynamics of the laser {\em subject to continuous observation}. This will be an important theme in my discussion of the thermodynamic limits to quantum clocks.

First however I would like to introduce an equivalent classical model. This can be found by computing the dynamics of the average field $\alpha(t)={\rm tr}(a\rho(t))$. Substituting this into the master equation and factorising moments (equivalent to neglecting quantum noise) the semiclassical dynamics is given by 
\begin{equation}
    \dot{\alpha}=-\frac{\kappa\alpha}{2}\left (1-\frac{Gn_s}{\kappa(|\alpha|^2+n_s)}\right )
\end{equation}
Well above threshold this is similar to the van der Pol oscillator in a rotating frame. There are two fixed points $\alpha_0=0$ and $|\alpha_0|^2= Gn_s/\kappa$ for $G >>\kappa$. The second solution is the above threshold limit-cycle solution.

To make a start  it is useful to find the stationary two-time correlation function
\begin{equation}
    g^{(1)}(\tau) = \frac{\langle a^\dagger(t+\tau) a(t)\rangle_{ss}}{\langle a^\dagger a \rangle_{ss}}
\end{equation}
Wiseman shows\cite{Wiseman99} that, well above threshold, this is given by
\begin{equation}
    g^{(1)}(\tau) = e^{-\kappa \tau/4\bar{n}}
\end{equation}
We can interpret this result as due to quantum phase diffusion noise at the rate $\Gamma= \kappa/2\bar{n}$. Note that this residual phase diffusion occurs at zero temperature unlike the classical case which goes to zero at zero temperature.  It suggests an ineliminable source of quantum noise for periodic clocks.

Wiseman\cite{Wiseman1993} presented the theory of a laser subject to continuous heterodyne detection. In the appropriate limit this is a continuous measurement of the real component of the electric field in the cavity in a frame rotating at the cavity frequency. Measurement changes the dynamics of the laser from the standard master equation given in Eq.(\ref{laser-me}). The reason for this is that measurement provides information about the state of the system and thus the conditional state of the system, conditioned on the measurement result must be considered. In other words, quantum state reduction must be modelled explicitly. 

To motivate this, suppose that at $t=0$, the state of the cavity field was projected by some hypothetical fast measurement into a coherent state $|\alpha\rangle$\cite{MilWalls}. This is a state of well defined phase and amplitude.  It is  clearly not a steady state of the system. How does it subsequently evolve according to Eq. (\ref{laser-me})? Following Wiseman\cite{Wiseman1993} we can solve the master equation for a short time interval $dt$ in the Fock basis to obtain
\begin{equation}
    \rho_{n,m}(dt)=e^{-|\alpha|^2}\frac{\alpha^n\alpha^{*\ m}}{\sqrt{n!m!}}\left (1-\frac{dt(n-m)^2}{2(n+m)}\right )
\end{equation}
where $\rho_{n,m}=\langle n|\rho|m\rangle$.  We ll above threshold we can replace $n+m$ in the denominator by $2\bar{n}$. This is the quantum analogue of linearising the noise on the classical limit cycle. It then follows that the short time dynamics is equivalent to that given by the master equation
\begin{equation}
\label{phase-diff-me}
    \dot{\rho}=\Gamma{\cal D}[a^\dagger a]\rho
\end{equation}
where $ \Gamma=\kappa/2\bar{n}$, the well know result for the line width of a laser. Eq. (\ref{phase-diff-me}) is the standard master equation description of phase diffusion. It ensures that the mean photon  number does  not change while the mean amplitude decays at the rate $\Gamma$. This corresponds to a state that is slowly diffusing in phase but all the while remaining of fixed intensity, completely analogous to phase difussion on a limit cycle in the classical case. 

In the case of continuous-time measurement we seek an equation that gives the conditional state in an infintesimal time interval $dt$ conditioned on all  previous measurement results parameterised by time.  This has come to be known as a quantum trajectory\cite{WiseMilb}. The description has two components: a classical stochastic differential equation to describe the fluctuating classical measurement record and a stochastic master equation to describe the corresponding change in the conditional state at each time step. 

In the case of optical heterodyne detection the classical stochastic differential equation describes a photo current record that results when the output of the laser cavity is mixed with a strong coherent local oscillator, detuned from cavity frequency, and then subject to photodetection\cite{WiseMilb}.  This current, suitably normlised, is given by 
\begin{equation}
    J_{het}(t)dt= \kappa \langle a(t)\rangle_cdt+ \sqrt{\frac{\kappa}{\eta}}dZ(t)
\end{equation}
where $0<\eta\leq 1$ is the photo-detector quantum efficiency, $dZ(t)$ is a complex valued Weiner process and $\langle a(t)\rangle_c$ is the usual trace average computed with respect to the conditional state $\rho_c(t)$ conditioned on the heterodyne current record to time $t$. This obeys the conditional master equation
\begin{equation}
   d\rho_c={\cal L}\rho_c dt+\sqrt{\frac{\eta}{\kappa}} {\cal H}[a]\rho_c dZ^*
\end{equation}
where the quantum conditioning superoperator ${\cal H}[A]$ is defined by 
\begin{equation}
    {\cal H[A]}\rho = A\rho+\rho A -{\rm tr}[A\rho +\rho A]\rho
\end{equation}
and ${\cal L}\rho$ is the rest of the master equation given in Eq.(\ref{laser-me}).
In essence this superoperator  partially `collapses' the quantum state at each time step. Note that if the system is in a coherent state, for which $a|\alpha\rangle=\alpha|\alpha\rangle$ at any time $t$, then the noise vanishes as ${\cal H}[A]|\alpha\rangle\langle \alpha|=0$. The conditional master equation tends to localise the conditional states on coherent states. In other words the continuous monitoring of the complex amplitude of the laser field tends to drive  the conditional states towards  a coherent state. 

Well above threshold when operating on the classical limit cycle we may use the phase diffusion master equation in Eq. (\ref{phase-diff-me}) and describe the conditional state of the system by
\begin{equation}
    \dot{\rho}_c= \Gamma {\cal D}[a^\dagger a]+\sqrt{\eta}{\cal H}[a]\rho_c dW^*
\end{equation}
Writing $\rho_c(t)$ as  a mixture of coherent states as
\begin{equation}
    \rho_c(t)=\int d^2\alpha P_c(\alpha,t)|\alpha\rangle\langle\alpha |
\end{equation}
and changing variables to $\alpha=r e^{i\phi}$ Wiseman\cite{Wiseman1993}shows that in the limit that $\eta\rightarrow 1$ the conditional phase dynamics well above threshold when the laser is operating on a limit cycle is stochastic and determined by
\begin{equation}
    d\phi_c=\sqrt{\Gamma}dW
\end{equation}
the standard phase diffusion model seen for a classical limit cycle in the presence of thermal noise. Here however the noise is not thermal in origin but quantum, ultimately due to spontaneous emission from the cavity in this model. The limit-cycle like behaviour emerges only when we consider the conditional state of a laser when the amplitude of the field is continuously monitored. 

The frequency of the field on the limit cycle in the original laboratory frame is close to the optical frequency of the cavity. This is very high which makes it impossible to count directly via standard electronic measures. Figuring out ways to do this is the core of atomic clock technology.  A first step can be made by including a non linear optical process into the laser cavity. In real optical clocks this is done by using a multi mode treatment to include many longitudinal cavity modes and incorporating an optical non linear process to couple them thus producing a mode locked laser.  In the single mode model presented here it suffices to include a single mode optical non linearity such as an atomic absorber or a Kerr non linearity. I will adopt the latter approach.

\subsection{Self-Modulated laser clocks}

\begin{figure}[ht]
\centering
\includegraphics[scale=0.45]{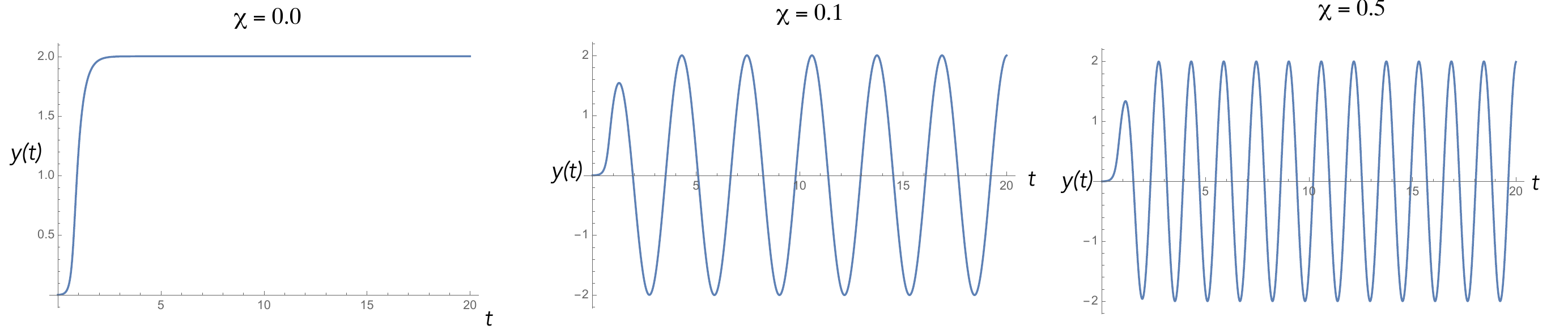} 
\caption{ Self-pulsing in a laser with an intracavity Kerr non linear medium. The imaginary part of the complex field amplitude for different values of the Kerr non linearity. The parameters are $\epsilon= 0.01,\delta=0.0, \kappa = 4.0, G=100, n_s=1.0$}
\label{laser-with-kerr}
\end{figure}

A Kerr optical non linearity is a material with an intensity dependent refractive index.  In a single mode model this would induce a time dependent detuning that depended on the number of photons in the cavity. With this in mind consider Eq.(\ref{laser-me}) changed to  
\begin{equation}
\label{laser-kerr}
  \dot{\rho}  =-i\epsilon[a+a^\dagger,\rho]-i\delta[a^\dagger a,\rho]-i\chi [a^{\dagger 2} a^2,\rho]+Gn_s{\cal D}[a^\dagger]\left ({\cal A}[a^\dagger]+n_s\right )^{-1}\rho +\kappa{\cal D}[a]\rho
\end{equation}
where $\epsilon$  is the amplitude of a small noisy injected signal at a carrier frequency $\delta$ detuned from the cavity frequency. The nonlinearity is paramterised by $\chi$ in units of frequency. 
The corresponding semiclassical equation is 
\begin{eqnarray}
    \dot{\alpha}  & =  & -i\epsilon-i(\delta+ 2\chi|\alpha|^2)\alpha\\\nonumber
    & & \mbox{              }-\frac{\kappa\alpha}{2}\left (1-\frac{Gn_s}{\kappa(|\alpha|^2+n_s)}\right )
\end{eqnarray}
The inclusion of a small driving term $\epsilon$ is important: without it the solutions to the semiclassical equations remain on the unstable fixed point at the origin in the classical limit without noise. It need only be a complex valued  Gaussian noise process with zero mean. It is unnecessary to include this in the full quantum model as quantum fluctuations suffice to drive the system.  The second term contains the intensity dependent detuning arising from the Kerr non linearity. In Fig. (\ref{laser-with-kerr}) I plot some examples of the self-pulsing limit cycle that develops from the original above-threshold laser limit cycle. 

A good estimate of the frequency of these oscillations can be found by linearising the dynamics on the  limit cycle. Using $\alpha=re^{i\theta}$ we find that the frequency on the limit cycle (for $\delta=0$) is given by $\Omega=\chi \bar{n}$.
The model so far is entirely deterministic and does not include the phase diffusion noise characteristic of a laser operating above threshold. A simple way to do this is to add the required phase diffusion noise for an above threshold laser to the semi classical phase equation to get the Ito equation (with $\delta=0$)
\begin{equation}
    d\theta =- \Omega dt +\sqrt{\Gamma}dW
\end{equation}
where $\Omega= \chi \bar{n}$ and $\Gamma =\kappa/2\bar{n}$. For $\bar{n}>>1$, the frequency becomes large and the noise becomes small.  As previously we define the (dimensionless) energy as $E=r^2$. On the limit cycle the energy is $E_*=\bar{n}$. The rate of change of energy is then given by 
\begin{equation}
    \dot{E} =-\kappa E +Gn_s=-\kappa E +\kappa \bar{n}
\end{equation}
for $G>>\kappa$.
The first term represents the rate of energy lost to the cavity via photon emission while the second term is the rate of energy gain due to the pumping mechanism.  The limit cycle forms when these two terms cancel. For a good clock we need the rate of phase diffusion to be small which requires  $\bar{n} >> 1$ in which case the energy dissipation is large: a good clock requires a large rate of energy dissipation.

\subsection{A quantum nano mechanical clock.}
In \cite{Park}  Park et al. demonstrated a nanomechanical oscillator driven by an electron tunnelling current. This system is an example of a periodic clock driven by a quantum noise process as I now show.  In the experiment a single $C_{60}$ molecule was loosely bound by van der Waals forces  between two gold electrodes. The frequency of the mechanical oscillation was $1.2$THz.  The mechanical motion can be approximated by a harmonic oscillator. An experimental implementation of a nanomechanical clock was recently reported by Pearson et al.\cite{Pearson2020} 

Single electrons tunnel on and off the molecule due to a tunnel coupling between a single quasi bound state on the molecule and fermionic reservoirs in the gold electrodes. These reservoirs are biased by a source-drain voltage and thus each is only in local thermodynamic equilibrium with a difference in their chemical potentials determined by the source-drain voltage. The molecule functions as a single electron quantum dot. When an electron tunnels onto the molecule it experiences an electric field that pushes it from the source to the drain whereupon the electron tunnels off the dot into the drain and the van der Waals potential provides a restoring force. In the right regime the molecule shuttles backwards and forwards on a limit cycle between the source and the drain electrode. The  nonlinearity responsible for the limit cycle is the exponential dependence of the tunnelling rate on the displacement of the molecule from eqilibrium.\begin{figure}[ht]
\centering
\includegraphics[scale=0.45]{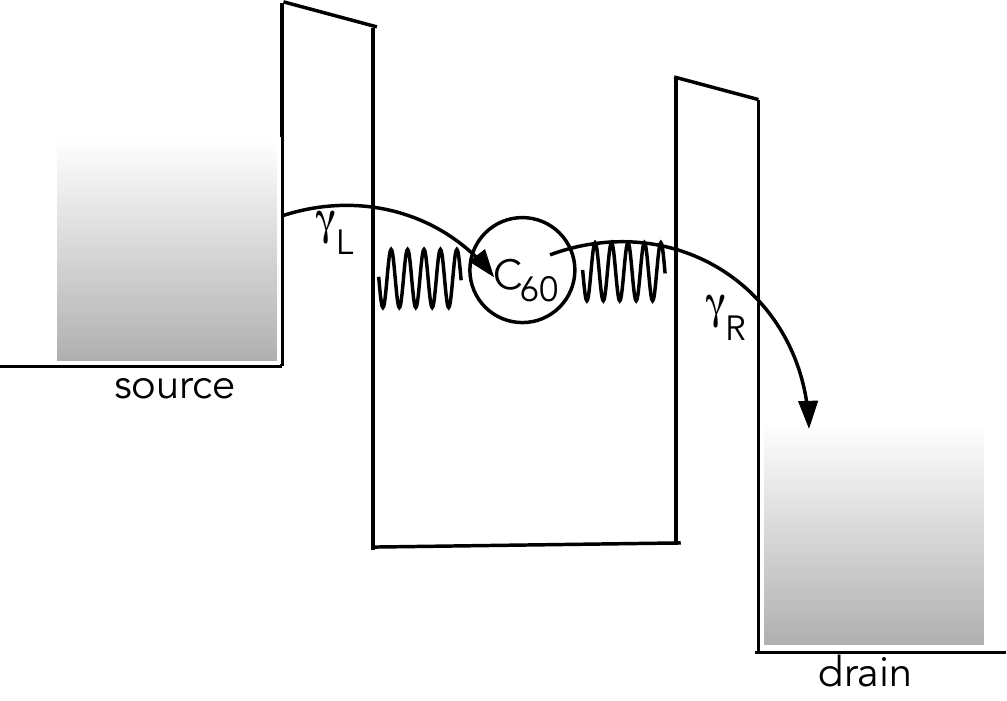} 
\caption{A schematic representation of a single harmonically bound quantum dot between two electrodes. Electrons tunnel onto the dot from the left and off the dot to the right. The bare tunnelling rates $]\gamma_L,\gamma_R$ are exponentially modulated by the displacement of the dot from equilibrium. }
\label{shuttle}
\end{figure}   A schematic diagram of this is shown in Fig.(\ref{shuttle}).

In \cite{Dian} a master equation treatment is presented and I will summarise that here. I will assume the so-called `zero temperature" limit in which, under appropriate bias conditions and very low temperature, the Fermi factor on the left is unity and the Fermi factor on the right is zero. Furthermore I will assume the bias conditions are such that the quasi bound electronic state on the molecule has an energy well below the Fermi level in the source and well above the Fermi level in the drain. This ensures that  low temperature fluctuations in the occupation close to the Fermi levels are not important.  

When electrons tunnel out of the source and onto the drain the external circuit increments/decrements the charge in the source/drain to ensure that they each remain in local thermodynamic equilibrium, i.e. keep a constant chemical potential. This links fundamental quantum discrete tunnelling events  to fluctuations of the current and voltages in the external circuit. 

The dynamical variables are the mean number of electrons on the molecule $\bar{n}$ and the mean displacement $X$ and mean conjugate momentum $P$ of the oscillator. The semiclasscial equations are obtained by calculating the equations of motion for these first order moments, using the master equation, and factorising all higher order moments;
\begin{eqnarray}
\frac{d \bar{n} }{dt} &=& \gamma_L (1 -\bar{n})e^{-4\eta X}- \gamma_R  \bar{n} e^{4\eta X} \label{e:shuttle-dyn1} \\
\frac{d\alpha}{dt} & = & -i\nu\alpha-\frac{\kappa}{2}\alpha+i\chi
\bar{n}\label{e:shuttle-dyn2}
\end{eqnarray}
with
$$
\alpha=\langle a\rangle=\langle \hat{x}\rangle/(2\lambda\eta)+i\langle\hat{p}\rangle\lambda\eta/\hbar\equiv X+iY\ \ .
$$
with $\nu$ the mechanical frequency of the molecular motion and
\begin{eqnarray}
\chi & = & eE\eta\lambda/\hbar\\
\eta & = & \left (\frac{\hbar}{2m\nu}\right )^{1/2}\frac{1}{\lambda}
\end{eqnarray}
while $\lambda$ sets the scale for the spatial dependence of the tunnelling rates on the molecular displacement from equilibrium, $e$ is the electronic charge and $E$ is the applied static electric field across the source-drain junction and $\kappa$ is the rate of energy dissipation in the mechanical oscillator.   
The system of equations, Eq.(\ref{e:shuttle-dyn1},
\ref{e:shuttle-dyn2}) has a fixed point, which undergoes a Hopf bifurcation with increasing $\chi$
 to a limit cycle forming the clock. The details can be found in \cite{Dian}.  Note the similarity between these equations and the equations of motion for the quartz oscillator coupled to a Schmidtt trigger given in Eq. (\ref{quartz-lc}) with  $g(V)=2V-1$. 
  \begin{figure}[ht]
\centering
\includegraphics[scale=0.5]{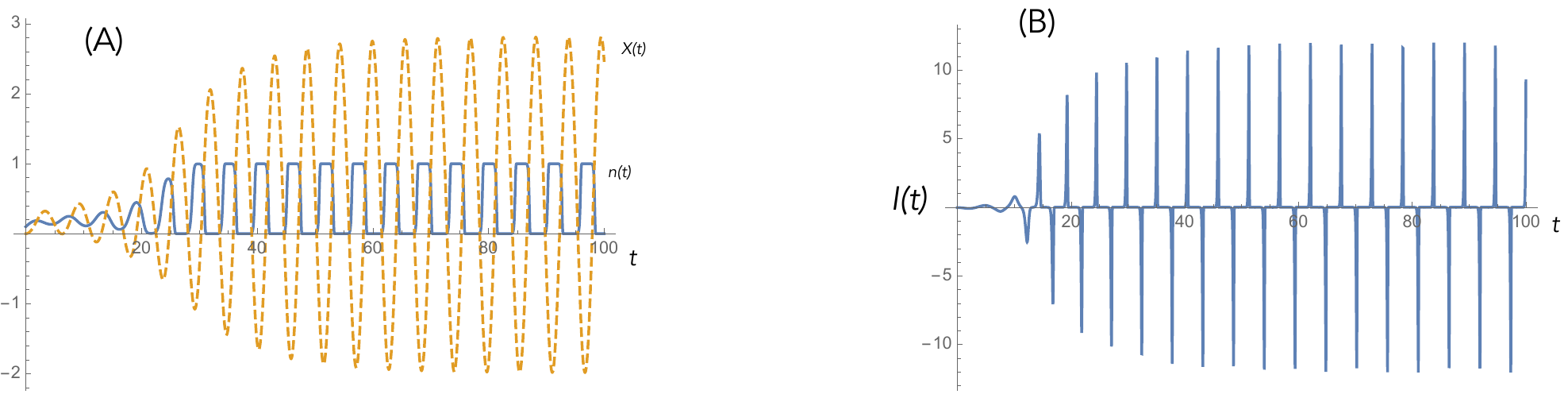}
\caption{ A typical limit cycle for the semiclassical model of a nanomechanical resonator coupled to tunnel junctions. (A) The charge, $n(t)$ and $x(t)=Re(\alpha(t)$ are plotted versus time. (B) The corresponding average current through the system, $2I(T)/e$.   Parameters are $\gamma_L=\gamma_R= = 0.1, \nu=1.0,\eta=1.0,\chi=1.0,\kappa=0.1$. } \label{shuttle-lc}
\end{figure}
 A typical limit cycle is shown in Fig.(\ref{shuttle-lc}). 
 
 A statistical description of the semiclassical dynamics can also be obtained from the master equation. The telegraph signal is determined by two point processes $dN_L(t)$ and $dN_R(t)$. When an electron tunnels onto the molecule in a time $dt$, $dN_L(t)=1$. Like wise when an electron on the molecule tunnels off in a time $dt$, $dN_R(t)=1$. The master equation determines the average values of these point processes in terms of quantum averages;
  \begin{eqnarray}
 E[dN_L(t)] & = & \gamma_L\langle e^{-4\eta \hat{X}} (\hat{n}-1)\rangle dt \\
  E[dN_R(t)] & = & \gamma_R\langle e^{4 \eta \hat{X}} \hat{n}\rangle dt
 \end{eqnarray}
  where $\hat{X}$ is the displacement operator for the mechanical motion and $\hat{n}$ is the number operator for the number of excess electrons on the molecule.
Correlations (entanglement) between  the electronic and motional states means these moments do not factorise in general. The conditions for which factorisation is a good approximation (see \cite{Dian}) define the semiclassical approximation. In that case the two Poisson processes become inhomogeneous point processes and the occupation probabilities obey
 \begin{equation}
     \dot{p}_1(t)=\gamma_Lp_0(t)e^{-4\eta X(t)}-\gamma_R p_1(t) e^{4\eta X(t)}
 \end{equation}
 The important point is that, unlike the apparently similar Quartz model Eq.(\ref{quartz-lc}), the noise in this case is not thermal diffusion in momentum but rather intrinsic quantum fluctuations arising from single electron tunnelling events. Nonetheless I will show that the limit for a good clock requires a large amount of energy dissipation.

 What is the clock signal? To answer that question we need to understand how the fundamental quantum tunnelling events are related to the current in the external circuit. These tunnelling events define conditional Poisson stochastic process $dN_L(T)$ and $dN_R(t)$ for left and right tunnelling events that  take the values $0,1$ in an increment of time $dt$, where $1$ corresponds to a single tunnelling event.  The rates for these process are conditioned by quantum averages. In a completely symmetric tunnelling scheme, $\gamma_L=\gamma_R=\gamma$, the average current is given by the classical stochastic process  
 \begin{equation}
 I(t)dt = \frac{e}{2}\left ( dN_L(t)+dN_R(t)\right )
 \end{equation}
It follows that in the semiclassical limit the average current is given by 
 \begin{equation}
   \bar{I}(t) = \frac{e\gamma}{2}\left (e^{-4\eta X} (n-1)+ e^{4 \eta X} n \right )
 \end{equation}
 An example is given in Fig(\ref{shuttle-lc}).
 
\begin{figure}[ht]
\centering
\includegraphics[scale=0.3]{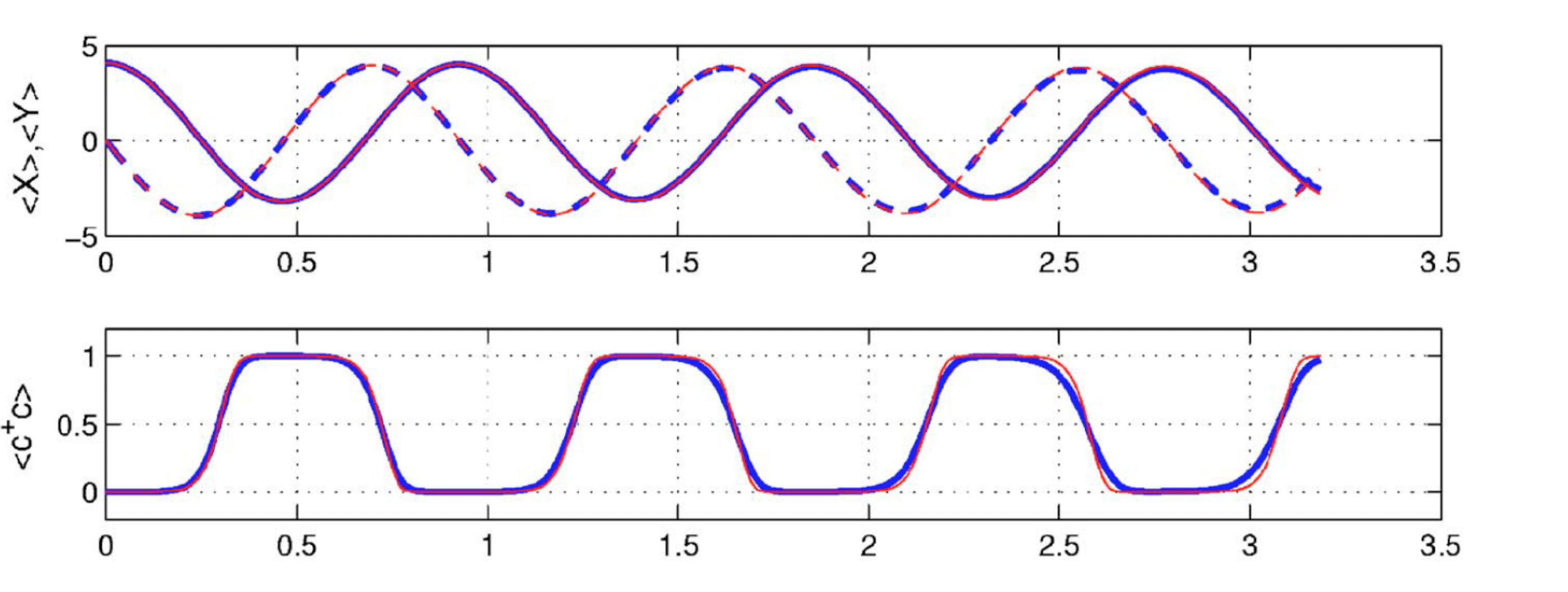}
\caption{The unconditional quantum dynamics of the nanomechanical oscillator in terms of the quantum averages $\langle c^\dagger c\rangle$,  and average displacement and momentum,$\langle X\rangle,  \langle Y\rangle $. Parameters are $\gamma_L=\gamma_R=1.0,\nu=1.0,\eta=0.3,\chi=`1.0,\kappa=0.05$  reproduced with permission from \cite{Dian}. } \label{unconditional-shuttle}
\end{figure}

The quantum dynamics is given by solving the master equation given in \cite{Dian}. If no attempt is made to monitor the current, the dynamics is given by the unconditional master equation. A typical example is shown in Fig.(\ref{unconditional-shuttle}).

\begin{figure}[ht]
\centering
\includegraphics[scale=0.3]{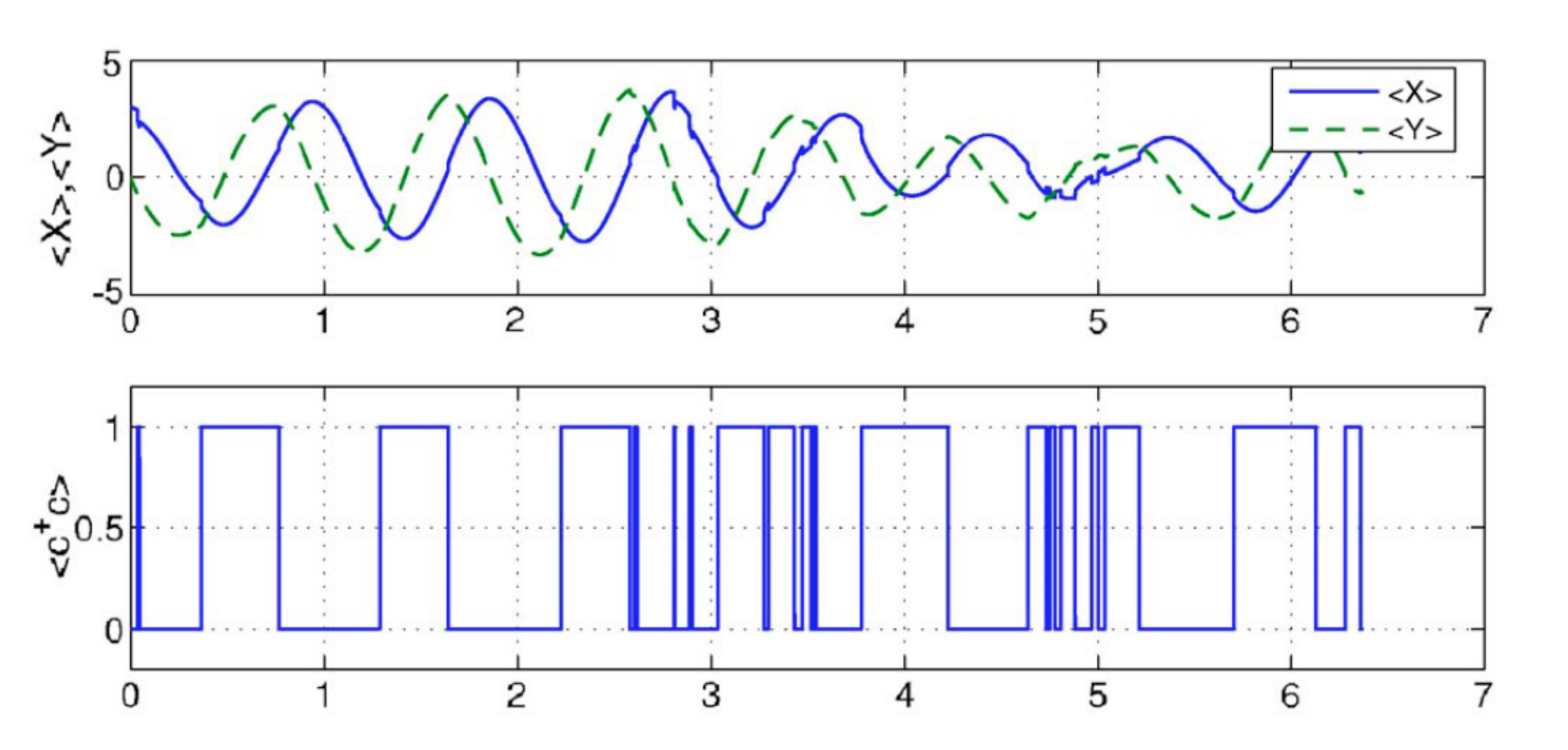}
\caption{Typical  conditional quantum dynamics of the nanomechanical oscillator in terms of the conditional quantum averages $\langle c^\dagger c\rangle$,  and average displacement and momentum,$\langle X\rangle,  \langle Y\rangle $. Parameters are $\gamma_L=\gamma_R=1.0,\nu=1.0,\eta=0.3,\chi=1.0,\kappa=0.05$  reproduced with permission from \cite{Dian}. } \label{conditional-shuttle}
\end{figure}

On the other hand if tunelling events are monitored using the current through the molecule, the dynamics is given by a conditional master equation. A typical example is shown in Fig.(\ref{conditional-shuttle}). 
The fluctuations in the clock period arising from the stochastic nature of quantum tunnelling events is now evident. This leads to phase and amplitude fluctuations in the dynamics of the mechanical oscillations limiting the accuracy of the clock. 

On the limit cycle we define $X(t)=r_*\sin(\Omega t), Y(t)=r_*\cos(\Omega t)$ where $r_*$ is the amplitude of the limit cycle and $\Omega$ is the frequency. Using the same two-time averaging procedure that I used in section (\ref{period-fluctuations}), to a good approximation $r_*$ is given by the solution to the transcendental equation
\begin{equation}
    e^{\mu x^2} = \cosh x
\end{equation}
where $x=\eta r_*$ and $\mu= \kappa\pi/(4\eta\chi)$. The frequency is well approximated by $\nu$ with small corrections of order $\nu^{-2}$.  

\begin{figure}[ht]
\centering
\includegraphics[scale=0.5]{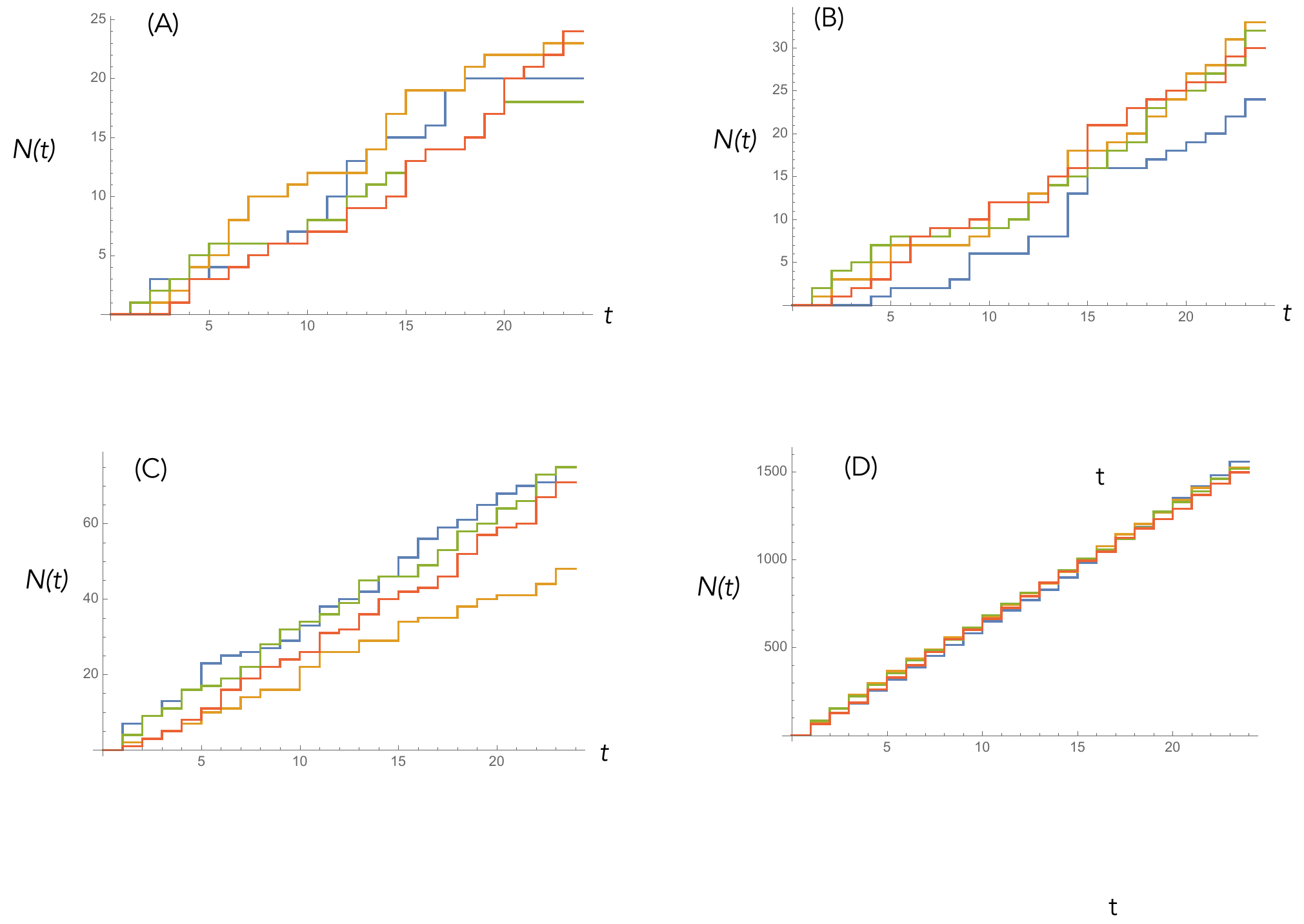}
\caption{A simulation of integrated count $N_R(t)$ for the process $dN_R(t)$, given that the molecule is occupied at $t=0$, for increasing values of the amplitude of the limit cycle $r_*$, (A)$r_*=0$, (B)$r_*=1.0$, (C)$r_*=2.0$, (D)$r_*=5.0$ and $\gamma_L=\gamma_R=1.0,\Omega=2\pi,\eta=0.3$ } \label{sample-dot}
\end{figure}

On the limit cycle the two inhomogeneous point processes describing the tunnelling events in the semiclassical limit are determined by the intensity functions
  \begin{eqnarray}
\lambda_L(t) & = & \gamma e^{-4\eta r_*\sin(\Omega t)}\\
\lambda_R(t) & = & \gamma e^{4 \eta r_*\sin(\Omega t)}
 \end{eqnarray}
The waiting time probability function for an electron to tunnell off the molecule is then given by
\begin{equation}
Pr(n(t)=1|n(0)=1)=\exp\left [-\int_0^t dt' \lambda_R(t')\right ]
\end{equation}
The regularity of the clock can be seen in the time derivative of the integrated count $N_R(t)$ for the process $dN_R(t)$. This is shown in Fig.(\ref{sample-dot}) for various values of $r_*$. In (A) the pure Poisson process with rate $\gamma$ is shown. This corresponds to $r_* =0$. As the size of the limit cycle increases the process becomes more regular. Similar results hold for $Pr(n(t)=0|n(0)=0)$. This means that in the limit of large limit cycles, the period of the conditional motion becomes increasingly regular and a good clock results. 

The thermodynamic constraints on this system are similar to the quartz clock with the important exception that here fluctuations are due entirely to quantum tunneling not thermal effects. As before the size of the limit cycle determines both the average energy dissipated and the quality of the clock. The semiclassical equations imply that the rate of change of the energy of the oscillator (in dimensionless units),  $E=(X^2+Y^2)/2$  is
\begin{equation}
    \dot{E}=-\kappa E(t)+\chi Y(t)n(t)
\end{equation}
 The first term on the right is energy lost due to damping into mechanical environment, this time at (near) zero temperature, while the second term is the electrical work done by the electrical field when the dot is occupied.   The argument runs through in a similar fashion to the case in section \ref{period-fluctuations}. In the limit of large limit cycle amplitude, $r_*$ large, the work done over one cycle, starting at the the time ($n(0)=1$), is given by the integral of $Y(t)$ over a single cycle which ends when the electron tunnels off the molecule at a time $t_{\downarrow}\Omega =\pi$ The work done on the $k'$ th cycle is just the integral 
 \begin{equation}
     W_k = \chi r_*\int_{0}^{t_{\downarrow}} dt \sin(\Omega t)=\frac{2\chi r_* T_k}{\pi}
 \end{equation}
 where $T_k=t_{\downarrow}$ and  $r_*$ is the radius of the limit cycle. Again this is in dimensionless units; to convert to SI units, multiply by $\hbar\nu$. As the work done on this cycle equals the energy  lost  due to damping $Q_k$ we find that 
 \begin{equation}
     Q_k=\frac{2\chi r_* T_k}{\pi\kappa }
 \end{equation}
 On average the energy dissipated depends on the average time the single electron occupies the molecule. Thus over a long time $\tau$, the average energy dissipated is determined by the average fraction of time the dot is occupied. Using standard results for a random telegraph signal we see that 
 \begin{equation}
     \bar{Q} = \frac{2\chi  r_*  \tau}{\pi\kappa }\left (\frac{\gamma_L}{\gamma_L+\gamma_R}\right )
 \end{equation}
This quantity is dimensionless (to convert to SI units multiply by $\hbar\nu$).  This example shows that, even at zero temperature, quantum fluctuations ensure that a good clock requires a large limit cycle $r_*$ and this corresponds to large energy dissipation in the molecular motion.

There is one final thermodynamic lesson to take away from this example. The system is ultimately driven by the different chemical potentials in the Fermi reservoirs in the source and drain.  The average energy that must be dissipated in the external circuit is given by $e V_{SD}(\mu_L-\mu_R)$ where $e$ is the electronic charge, $V_{SD}$ is the source drain bias voltage and $\mu_{L/R}$ are the chemical potentials in the left and right reservoirs. That the chemical potentials are the not the same of course indicates that the entire system is not in thermal equilibrium; the essential requirement for a clock to operate. The free energy has been increased by electrical work done on it by the external circuit. Erker et al.\cite{Erker2017} give an interesting example in which a clock is driven by a difference in temperature (the other Lagrange multiplier) between two reservoirs. There is another very important class of clock that is driven by a difference in chemical potential; the chemical clocks that characterise biological systems. Perceived time is determined by the irreversible processes in such clocks \cite{Buonomano}.

\section{Non periodic clocks: information as fuel.}
\label{non-periodic-clocks}

At first sight the notion of a non periodic, irreversible clock seems paradoxical:  a clock is the very epitome of a deterministic periodic system.  Yet irreversible clocks have long been used, water clocks for  example, and a particular irreversible clock is in common usage.  Radiocarbon dating is based on the stochastic and irreversible decay of a radionuclide---$\text{C}^{14}$. Metabolism ensures living organisms will continually refresh their concentration of $\text{C}^{14}$ while alive, but once dead the concentration begins to relax to the steady state non-organic concentration of $\text{C}^{14}$ in the environment.  Indeed, in radiocarbon dating one must take into account that different organisms die with differing concentrations of $\text{C}^{14}$. Furthermore, the vital concentration may change with time due to concentration of $\text{C}^{14}$ in the atmosphere changing over time. In other words, different organisms and different epochs do not necessarily begin with the same non equilibrium distribution. As we will see, relaxation from a non equilibrium state is a key feature of non deterministic clocks, especially when the non equilibrium state is a result of state conditioning of a system subject to measurement.

\subsection{Radio carbon clock.}
 It is known that the probability of a single radionuclide not to decay in a time $t$ is $\exp(- \gamma t)$, where $\gamma$ is called the decay rate.  Given an ensemble of undecayed radionuclides, the number that have decayed after some time $t$ is a Poisson distributed stochastic variable $N$ with mean $\gamma t$.  Thus, given a count of the number of radionuclides that have decayed, $N$, we may estimate the time that has elapsed as
\begin{equation}
	t_{\text{est.}} = \frac{N}{\gamma} .\label{eq:radiocarbonEstimate}
\end{equation}
The error in this estimate is given by the fluctuations in the final count $\delta N$ for a fixed estimate $t_{\text{est.}}$.  Since for a Poisson process the variance is equal to the mean, one finds the relative error is
\begin{equation}
	\frac{\delta t_{\text{est.}}}{t_{\text{est.}}} = \frac{1}{\sqrt{\gamma t_{\text{est.}}}} .\label{eq:radiocabonError}
\end{equation}
Conveniently, we get a better estimate the longer we wait.  Now, in the case of radio-carbon dating, one indeed wishes to compute the elapsed time $t_{\text{est.}}$, and so one requires to know $\gamma$ before hand; calibrate the clock.  However, conversely, one could instead take this system as defining a temporal scale, where to take the time would be to make the count $N$, and the value of $\gamma$ would merely reflect a particular choice of units and thus be purely conventional.  That is, the actual count $N$, while subject to fluctuations, is a local physical quantity that can serve as physical time.  Yet it is worth noting that a radionuclide can only decay, i.e., $N$ can only increase, so in contrast to the prototypical reversible clock, this is a clock with no tock.

\subsection{Mach's thermal clock}

\begin{figure}[ht!]
\centering
\includegraphics[scale=0.5]{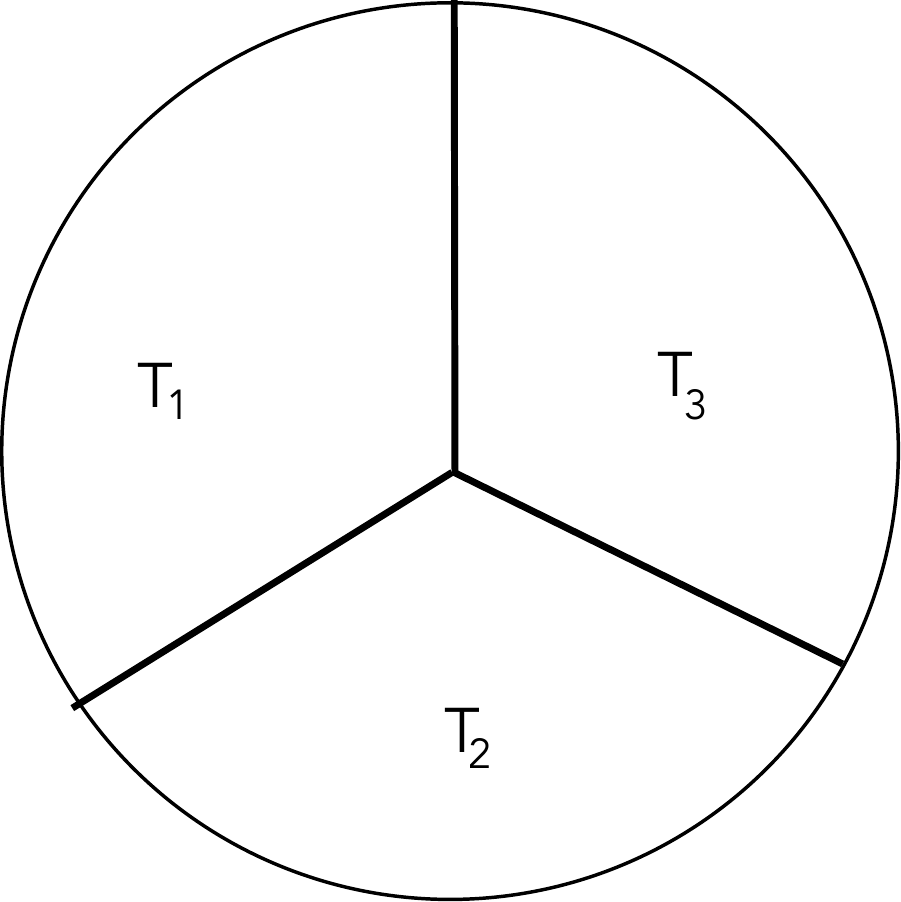} 
\caption{A schematic diagram of Mach's hypothetical thermal clock. Three identical materials, initially at different temperatures, are placed in thermal contact but completely isolated from the rest of the universe.  By monitoring the temperature of any one of them, elapsed time can be estimated. }
\label{mach-clock}
\end{figure}

  Mach's thermal clock was an attempt to ground an understanding of time in our sense perception of irreversible processes~\cite{Mach}.   Mach's clock is a thermodynamic system in which three identical materials, initially held at different temperatures, are placed in thermal contact but thermally isolated from the rest of the universe, see Fig.(\ref{mach-clock}). 

As the system is isolated, the internal energy, $U$ cannot change with time, thus 
\begin{equation}
\dot{U} =c(\dot{T}_1+\dot{T}_2+\dot{T}_3)=0
\end{equation}
where $c$ is the specific heat. 
Thus $T_1(t)+T_2(t)+T_3(t) = T_1(0)+T_2(0)+T_3(0) $, a constant. Newton's law of cooling, is an empirical law that says the rate of change of temperature is proportional to the temperature difference.  It is used by forensic pathologists. In this case it implies that 
\begin{equation}
    \dot{T}_1 = k(T_3-T_1)+k(T_2-T_1)
\end{equation}
which is clearly negative and $k$ is a constant.  Using similar relations for the other temperatures we find that 
\begin{equation}
\label{mach-time}
\bar{T}-T_1(\tau)=A e^{-3k\tau}
\end{equation}
where $A=\bar{T}-T_1(0)$ is a constant and we have defined the initial average temperature as $\bar{T}=(T_1(0)+T_2(0)+T_3(0))/3$.  This is very like radioactive decay.  Given a good thermometer, it is possible to estimate the time and conversely.
Eq. (\ref{mach-time})  implies that the uncertainties satisfy
\begin{equation}
\label{mach-tolman}
    \frac{\Delta T_1}{T_1}+\frac{\Delta \tau}{\tau_{est}}=0
\end{equation}
where I have defined an estimated time by $\tau_{est}=1/3k$ in analogy with radioactive decay.


\subsection{Quantum thermal clocks.}
\label{TLS-clock}
A simple generalisation that introduces temperature in a fundamental way results if we consider a two-level system (TLS) in a thermal bath at temperature $T$ (radiative damping).  While the ensemble average is a stationary state in equilibrium with the bath, continuous measurement of the energy of a single realisation (discussed in more detail below) yields a random telegraph signal.  Let us define the random telegraph signal $Z(t)$ to take the value $Z(t) = -1$ if the two-level system is found in the ground state $|g\rangle$ and $Z(t) = 1$ if it is found in the excited state $|e\rangle$.  The conditional state at time t, given the measurement record up to that time, is then
\begin{equation}
	\rho_{\text{cond.}}(t) = \frac{1 + Z(t)}{2} |e\rangle \langle e| + \frac{1 - Z(t)}{2} |g\rangle \langle g| .
\end{equation}
On the other hand, the ensemble average---the average over all possible random telegraph signals---yields the thermal state
\begin{equation}
	\rho_{\text{th.}}(\beta) = \frac{1 - \tanh(\beta \epsilon / 2)}{2} |e\rangle \langle e| + \frac{1 + \tanh(\beta \epsilon / 2)}{2} |g\rangle \langle g| \label{eq:thermalState}
\end{equation}
where $\beta = 1/k_BT$ with $T$ the  temperature of the bath, $k_B$ is Boltzmann's constant and $\epsilon$ is the energy separation between $|g\rangle$ and $|e\rangle$.  (The convention is $|g\rangle$ is assigned energy $- \epsilon / 2$ and $|e\rangle$ is assigned energy $\epsilon / 2$.)  The transition rate from $|e\rangle$ to $|g\rangle$ is $\gamma (\bar{n} + 1)$ and that from $|g\rangle$ to $|e\rangle$ is $\gamma \bar{n}$ where $\bar{n} = [\exp(\beta \epsilon) - 1]^{-1}$ is the bath occupation and $\gamma$ is the spontaneous emission rate.  At temperatures and energies such that $\bar{n} >>1$ we can neglect spontaneous emission and these rates are the same.

A simple irreversible clock can now be defined operationally by counting the number of transitions $N$.  As before, $N$ is a Poisson distributed stochastic variable, but this time with parameter $\gamma\bar{n}$, which depends on temperature.  In the high-temperature limit, the equivalent to Eq.~\eqref{eq:radiocarbonEstimate} reads
\begin{equation}
	t_{\text{est.}} =\dfrac{N}{\gamma\bar{n}} \approx\frac{\epsilon N}{\gamma k_BT} .
\end{equation}
The error in the estimate scales as $(\gamma\bar{n})^{-1/2}$ which is the rate at which energy is dissipated. As is the case for periodic clocks,  the greater the rate of energy dissipated the better the clock.  This is the key result of Erker et al. \cite{Erker2017}.

In contrast to the example from radiocarbon dating, here the relative error in $t_{\text{est.}}$ is limited by the relative error in the temperature, $T$.  That is, in order for this system to define an accurate temporal scale, we require a good thermometer.  Conversely, we may use this system as a thermometer where the temperature is estimated from elapsed time,
\begin{equation}
	k_BT_{\text{est.}} = \frac{\epsilon N}{\gamma t} ,
\end{equation}
so one may equally well say that in order for this system to define an accurate temperature scale, we require a good clock.  For a fixed count $N$, and a given two-level system, we see that for high temperature, 
\begin{equation}
\label{tolman-TLS}
  t_{est}.T_{est}  = constant
\end{equation}
which implies that
\begin{equation}
	\frac{\delta t_{\text{est.}}}{t_{\text{est.}}} + \frac{\delta T_{\text{est.}}}{T_{\text{est.}}} = 0 .
\end{equation}
This is the same structure that was derived for the Mach clock in Eq. (\ref{mach-tolman}) and suggests a special relationship between time and temperature that I will return to in section sections (\ref{tolmanGR},\ref{rovelli-thermal-time}). An elegant  treatment of the statistical connection between time and temperature has recently been given by Tanaka\cite{Tanaka} using the Fisher information metric associated with estimation of time and temperature.  

There is a hidden assumption in this treatment: I have  assumed that one can indeed monitor the state of a single TLS and thus accurately sample the random telegraph process $Z(t)$. In this assumption I have in effect smuggled in a very low entropy bath to ensure the measurements do indeed provide reliable information on the state of the TLS. In this type of clock, the conditional state of the system is pushed away from thermal equilibrium, and the free energy raised, not by doing work but by extracting entropy through good measurements. In this sense, information is fuel.  

\subsection{A superconducting quantum thermal clock.}
\begin{figure}[ht]
\centering
\includegraphics[scale=0.45]{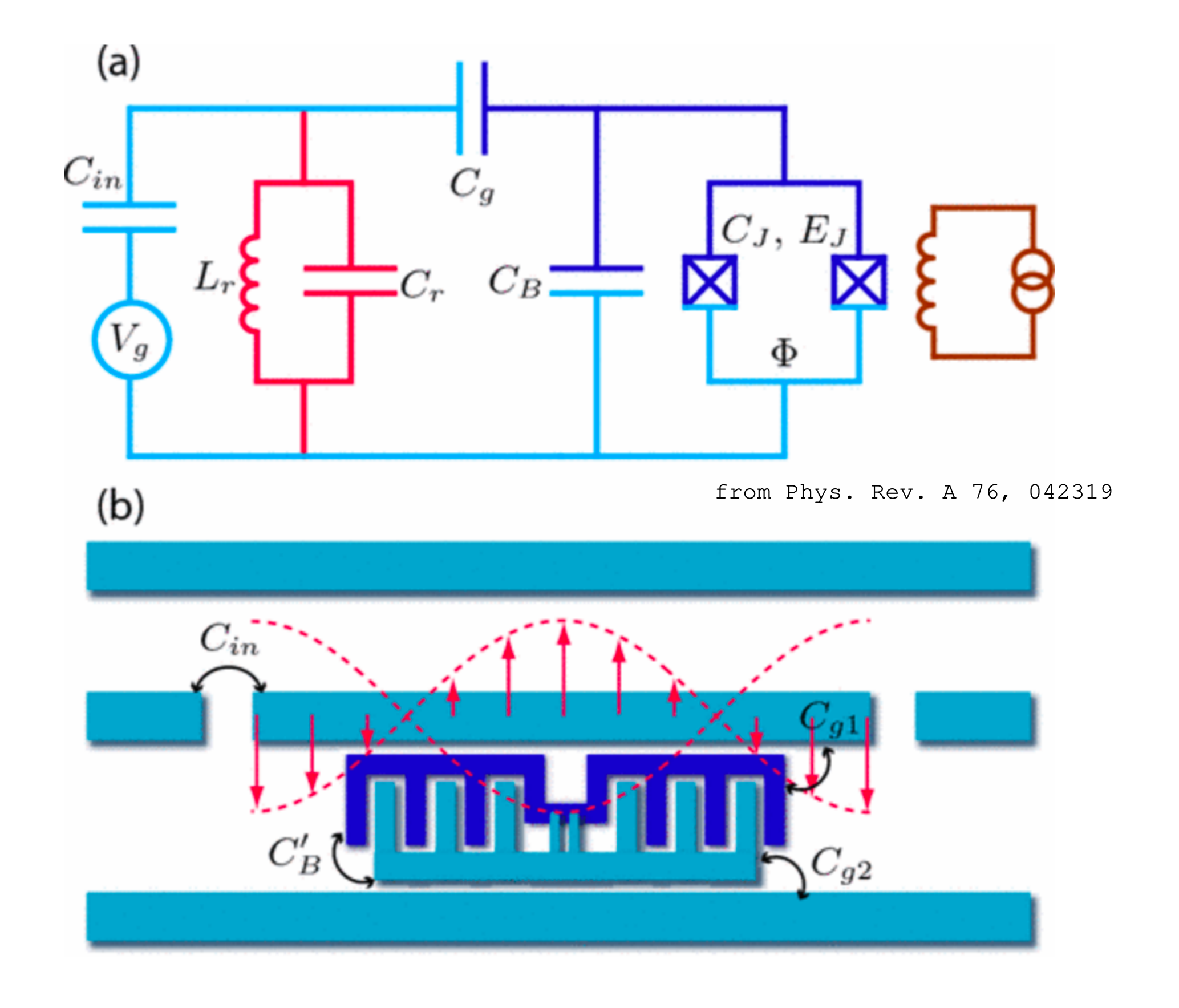} 
\caption{An equivalent circuit model for a superconducting transmon in a co planar cavity together with a schematic representation of the device. Reproduced with permission from Koch et al Phys. Rev. A 76, 042319\cite{transmon} }
\label{transmon-fig}
\end{figure}

It is  not easy to make detectors that count individual quantum jumps of a quantum system but it is a standard measurement for a TLS based on superconducting quantum circuits~\cite{Vijay2011}. These are devices based on the Josephson effect in superconductors at very low temperatures. A typical such device is known as a transmon\cite{transmon} depicted in a highly simplified schematic form in Fig.(\ref{transmon-fig}). 

A small metal superconducting island is separated from a reservoir of Cooper pairs by tunnelling junctions. Cooper pairs can tunnel on and off the island though the Josephson effect. Furthermore a hole is introduced into the junction forming a `split junction'. By changing the magnetic flux through this hole, the coupling energy between the transmon and the reservoir can be changed, effectively changing the  rate of the tunnelling. In addition, the island can be capacitively coupled to a neighbouring superconducting bias line enabling the Coulomb energy of the transmon to be changed.  Thus there are two `handles' for changing the energy of the transmon.  By carefully adjusting the ratio of the capacitive coupling to the tunnel coupling a discrete energy spectrum results. These levels are not equally spaced enabling a microwave voltage source on the capacitive coupling to be resonant only with transitions from the ground state to the first excited state. This is the two-level system of interest. The microwave voltage is sustained in  driven superconducting cavity containing the transmon. Effectively this constitutes a simple TLS in an electromagnetic cavity that may be thermally or coherently driven, or both. 

We now require a way to measure the state of the TLS. This is based on the method of dispersive coupling~\cite{Johnson}. The transmon can be simultaneously coupled to a second cavity with a resonant frequency that is well-detuned from the TLS transition frequency. In this case transitions between the energy eigenstates of the TLS are suppressed but the cavity frequency can be shifted by different amounts depending on the state of the TLS. This is a non linear dispersion. Reciprocally, the energy difference between the states of the TLS are shifted by an amount proportional to the intensity of the field in the cavity. In atomic physics this is known as the dynamic Stark effect.  We will refer to this dispersively coupled cavity as the {\em readout cavity}.  For a more detailed discussion see section 3.10.2 of \cite{WiseMilb}

If the readout cavity is thermally isolated and coherently (but weakly)  driven, the steady state field inside the cavity is in a coherent state: a pure state of zero entropy. The amplitude of this state depends on the detuning of the driving field from the readout cavity resonance frequency. But as this resonance frequency is shifted by the TLS, the phase of the  field emerging from the cavity will change depending on the state of the TLS. If this phase is continuously monitored using homodyne detection, a continuous measurement record of the changing state of the TLS can be inferred under the right conditions. One of these conditions is that the readout cavity must be subject to very little thermal excitation so that it remains close to a coherent state. This amounts to assuming that the readout system has very low entropy thus making  explicit the necessity for another very low temperature heat bath for the thermal clock to function. 

\begin{figure}[ht]
\centering
\includegraphics[scale=0.6]{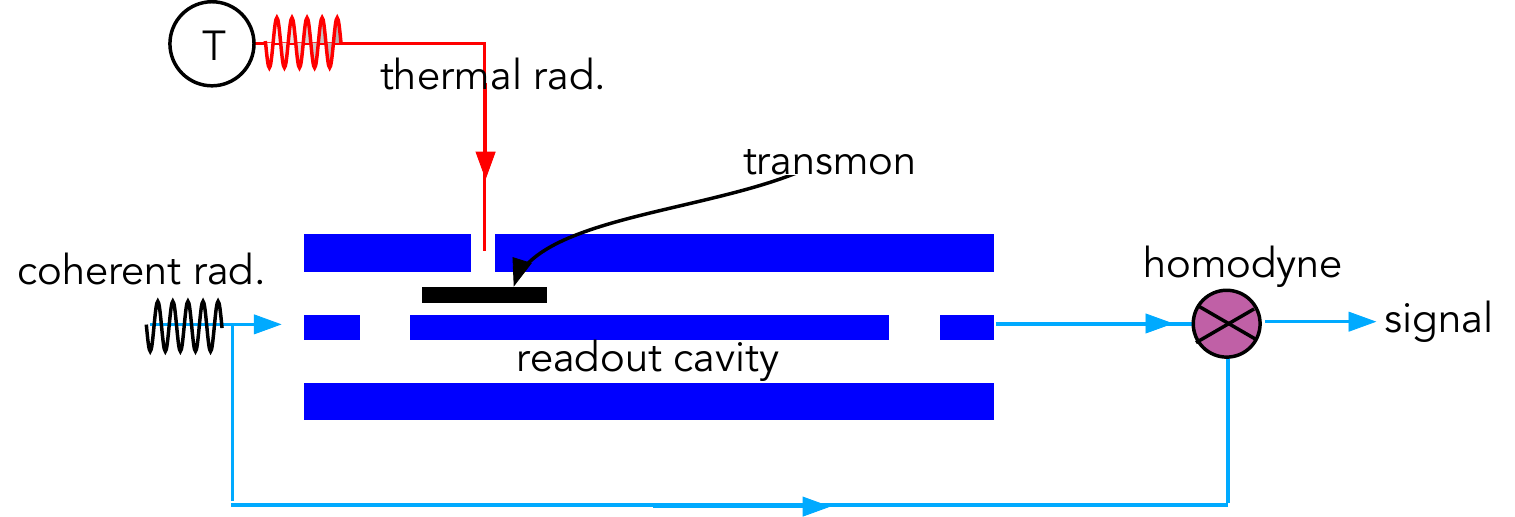} 
\caption{A quantum superconducting thermal clock. A transmon two-level system (TLS) is dispersively coupled to a coplanar waveguide cavity on a chip. A coherent microwave field is transmitted through the cavity and detected via homodyne mixing producing a signal that monitors the state of the TLS in real time. To produce a non periodic thermal clock the TLS is driven by high-temperature ($T$) noise, generated by an arbitrary wave generator and up-converted to the TLS frequency with the help of an external microwave source [Figure courtesy of Arkady Fedorov].
}
\label{scheme}
\end{figure}

To implement a thermal clock with this system, the TLS must be coupled to a thermal radiation field.  This can be simulated by driving the TLS with an additional Gaussian noise field at the TLS resonance to implement controllable thermal noise acting on the transmon.  
A schematic of this temperature clock  is given in  Fig.(\ref{scheme}).

The joint quantum state of the resonator and the readout qubit satisfies the master equation
\begin{eqnarray*}
\dot{\rho}  & = &  -iE[a+a^\dagger,\rho]-i\chi[a^\dagger a\sigma_z,\rho]\\
& & +\gamma(\Bar{N}+1){\cal D}[\sigma_-]\rho+\gamma\Bar{N}{\cal D}[\sigma_+]+\kappa{\cal D}[a]\rho,
\end{eqnarray*}
where $E$ is the amplitude of the coherent driving field driving the measurement cavity, $\chi$ is the strength of the dispersive coupling between TLS and cavity, $\kappa,\gamma$ are the energy decay rates of the readout cavity and TLS respectively, and $\bar{N}$ is the mean excitation in thermal radiation coupled to the TLS.  The final four terms describe irreversible evolution. The superoperator appearing in these terms is defined, as previously, by ${\cal D}[A]\rho=A\rho A^\dagger-\frac{1}{2} A^\dagger A\rho -\frac{1}{2}\rho A^\dagger A$.  

Note that the coupling of the TLS to the cavity is proportional to the energy of the TLS which ensures that this coupling, of itself, cannot change the energy distribution of the TLS but it does change the detuning of the measurement cavity, conditional on the state of the TLS, and thus changes the amplitude of the field emitted.  The homodyne current (scaled by $\gamma$) of the emitted field is given by 
\begin{equation}
I(t) =\langle a+a^\dagger\rangle_c +\xi(t)/\sqrt{\gamma\eta}
\end{equation}
where $\xi(t)$ is a white noise process, $0<\eta<1$ is the homodyne detection efficiency, $\gamma$ is the TLS energy decay rate.  This classical stochastic process will be used to monitor the quantum jumps of the TLS and thus constitutes the counter for the thermal clock.

While the current $I(t)$ is a classical stochastic process it's statistics is conditioned on the quantum source; thus the subscript on the quantum average in the first term.  The conditional state, conditioned on a particular current history can be computed using the method of quantum trajectories\cite{WiseMilb}. In effect this describes a continuous stochastic state reduction up to time $t$. 
This state obeys the stochastic master equation~\cite{WiseMilb}
\begin{eqnarray*}
d\rho_c  &= & -iE[a+a^\dagger,\rho_c]-i\chi[a^\dagger a\sigma_z,\rho_c]\\
 &&+\gamma(\Bar{N}+1){\cal D}[\sigma_-]\rho_c+\gamma\Bar{N}{\cal D}[\sigma_+]\rho_c
+\kappa{\cal D}[a]\rho_c+\sqrt{\kappa\eta}{\cal H}[a]\rho_c dW(t),
\end{eqnarray*}
where $dW(t)$ is a classical Wiener process and ${\cal H}[A]\rho=A\rho+\rho A^\dagger -{\rm tr}(A\rho+\rho A^\dagger)$.

To operate as a pure entropy clock, it is necessary to distinguish quantum jumps in the TLS due to thermal excitations. To this end we will assume that the readout cavity is strongly driven  and rapidly damped. In this limit the state of the TLS obeys the effective master equation in the Schr\"{o}dinger picture,
\begin{equation}
\dot{\rho}=-i(\Delta+\epsilon/\hbar)[\sigma_z,\rho]+\gamma(\Bar{N}+1){\cal D}[\sigma_-]\rho+\gamma\bar{N}{\cal D} [\sigma_-]\rho+\Gamma {\cal D}[\sigma_z]\rho,
\label{master-measurement}
\end{equation} 
where $\Delta = 4\chi E^2/\kappa^2$ and $\Gamma=4\Delta\chi/\kappa$. The last term in the master equation leads to pure phase-diffusion around the z-axis on the Bloch sphere and decay of the TLS polarisation but cannot change the steady-state of the TLS which remains thermal. It also describes a continuous measurement of $\sigma_z$ as I now explain. 

The observed record is the homodyne stochastic photocurrent exiting the readout cavity of the qubit that now becomes 
\begin{equation}
I(t)dt =\langle \sigma_z\rangle_c dt+dW(t)/\sqrt{\eta\Gamma},
\end{equation}
The conditional stochastic master equation now takes the form
\begin{eqnarray*}
d\rho_c & = &-i(\Delta+\epsilon/\hbar) [\sigma_z,\rho_c]dt+\gamma(\Bar{N}+1){\cal D}[\sigma_-]\rho_c dt+\gamma\bar{N}{\cal D} [\sigma_+]\rho_c dt\\
&&+\Gamma {\cal D}[\sigma_z]\rho_c dt+\sqrt{\eta\Gamma}{\cal H}[\sigma_z]\rho_c dW(t)
\end{eqnarray*}

The condition to see quantum jumps is $\Gamma >>\gamma\bar{N} $ which ensures that information about the energy of the TLS is acquired faster than it is changed by thermal transitions.
\begin{figure}[ht!]
\centering
\includegraphics[scale=0.5]{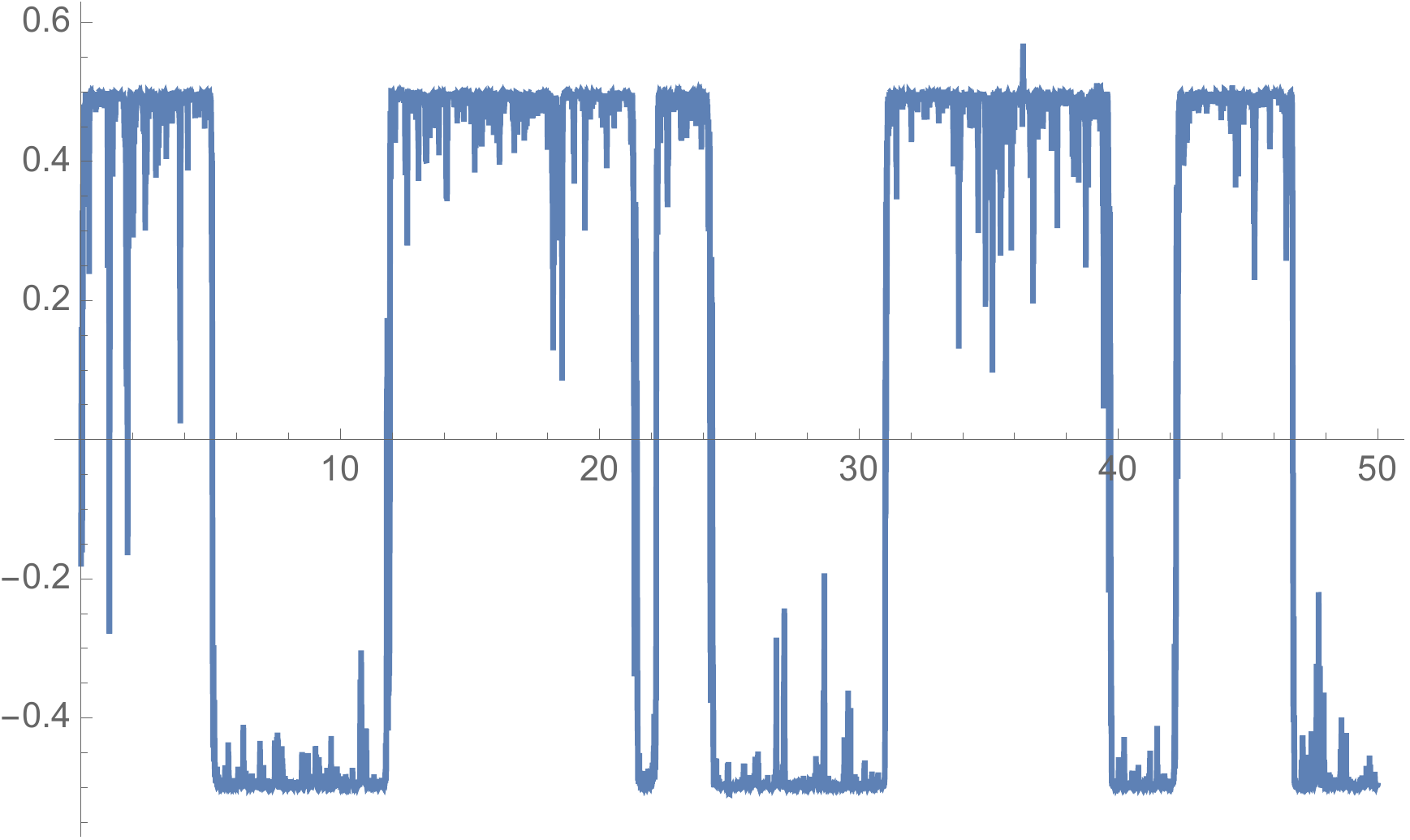} 
\caption{ The conditional mean energy of the TLS (in units of $\epsilon$) versus time in the good measurement limit. The jumps are caused by photons entering and exciting the TLS from the thermal bath coupled to it. The mean thermal photon number in the bath was $\bar{N}=0.9$ and $\Gamma=5.0$, $\gamma=0.1$}
\label{quantum-jumps}
\end{figure} 
In  Fig.(\ref{quantum-jumps}) I plot the conditional mean of $\sigma_z$ showing quantum jumps. In this limit the system is operating as a thermal clock driven by reductions in entropy due to the measurement. In other words, when the rate of information gain by an observer exceeds the rate of thermally driven fluctuations we obtain a thermal clock. In this limit the free energy of the clock is increased by entropy reduction rather than work done as in the typical periodic clock.

\section{Fundamental questions relating to clocks and thermodynamics.}

An understanding of the thermodynamics of clocks suggest novel responses to deep questions in fundamental physics and philosophy. Mach\cite{Mach} (see Fig (\ref{mach-clock})) and also Eddington\cite{Eddington} introduced the idea of a thermal or entropy clock to probe the question of the arrow of time. Eddington introduced the idea of an entropy clock in an attempt to penetrate the mystery of perceived time. As he put it \cite{Eddington}
\begin{quote}
    It seems to me, therefore, that consciousness with its insistence on time's arrow and its rather erratic ideas of time measurement may be guided by entropy-clocks in some portion of the brain. [\dots] Entropy-gradient is then the direct equivalent of the time of consciousness in both its aspects. Duration measured by physical clocks (time-like interval) is only remotely connected. 
\end{quote}
I think you might agree, having read this far,  that his conclusion is misguided: all physical clocks are necessarily irreversible. In this sense all clocks are entropy clocks. The link to consciousness is a more interesting speculation. 

Our understanding of time has been enriched by the General Theory of Relativity  but Eddington's comment seems to suggest that the time of GR is somehow different to the kind of thing measured by the clocks I have discussed in this paper. This is certainly not the case. The Tolman relations in gravitational thermodynamics points to a deep connection between the time of GR and the thermodynamics of clocks by showing a duality between proper time and temperature. In seeking a quantum theory of gravity the problem of time has led Rovelli\cite{Rovelli-book} to propose the `thermal time hypothesis'. Thermal clocks give operational significance to this. 
In the remainder of this paper I will comment on these issues in the light of a deeper understanding of the thermodynamics of clocks.

\subsection{General relativity and Tolman relations.}
\label{tolmanGR}
In his book {\em Relativity, Thermodynamics and Cosmology} \cite{Tolman}  Tolman applied the theory of general relativity to thermodynamical systems. The most significant result from the perspective of this paper is that for a gas in thermal equilibrium and  gravitational equilibrium the temperature cannot be uniform and must vary from place to place according to the metric of the relevant space time. 

Before presenting the general relativistic case I will begin with the special relativistic case.  Consider an observer at rest with respect to a two-level system operating as a thermal clock (see section \ref{TLS-clock} ) in contact with a heat bath with a temperature $T_0$ with respect to that observer. In this case the relation between the estimated elapsed proper time $\tau$ and $T_0$ was seen to be $T_0\tau = constant.$ (Eq.(\ref{tolman-TLS}). For an observer in uniform relative motion at relative velocity $v$, the elapsed time is dilated to be $t'=\tau/\sqrt{1-v^2/c^2}$. Given that the physics of the clock and its interaction with the heat bath are unchanged, we see that for the moving observer the temperature will be estimated to be $T'=T_0\sqrt{1-v^2/c^2}$; the correct transformation of temperature in special relativity ( see section 69 of \cite{Tolman}).

Tolman developed a general covariant definition of thermal equilibrium along quite general lines in \cite{Tolman}.  I will summarise this result here in Tolman's notation for the case of a sphere of gas held together by gravitational forces and in thermal equilibrium. In this case the general invariant interval can be written as 
\begin{equation}
    ds^2= -e^\mu(dr^2 +r^2d\theta^2+r^2 \sin^2\theta d\phi^2)+e^{\nu}dt^2
\end{equation}
where $r,\theta,\phi$ are standard spherical polar coordinates and $\mu=\mu(r), \nu=\nu(r)$. In this case Tolman's condition for thermal equilibrium reduces to 
\begin{equation}
    \frac{d}{dr}\left (e^{\mu/2} r^2\frac{d}{dr}\left (\frac{1}{T_0(r)}\right )\right ) =\frac{4\pi(\rho_{00}+3p_0)}{T_0(r)} e^{\mu/2} r^2
\end{equation}
at all values $r$ inside the sphere and 
where $T_0(r)$ is the proper temperature in a thin spherical shell at radius $r$, while $\rho_{00}$ and $p_0$ are the mass density and pressure in that shell. Integrating this equation 
\begin{equation}
    \frac{d\log T_0(r)}{dr}=-\frac{1}{2}\frac{\nu(r)}{dr}
\end{equation}
equivalently
\begin{equation}
    T_0\sqrt{g_{44}}=\mbox{constant}
    \label{Tolman}
\end{equation}
where $g_{44}=e^\nu$ is the temporal component of the metric. The same result holds for the case of a sphere of pure black-body radiation.  The result Eq.(\ref{Tolman}) also follows by using a two-level thermal clock to measure proper time at radius $r$, in thermal contact with the local environment at temperature $T_0$ (section  (\ref{TLS-clock})).

\subsection{Rovelli's thermal time hypothesis.}
\label{rovelli-thermal-time}
The variable $t$ that appears in the theory of general relativity, like the other coordinates, is arbitrary yet  when a physical clock is localised at a space time point it should count events in accordance with this coordinate. That does seem to make the $t$ coordinate special.  How do we identify such systems?  Rovelli\cite{Rovelli-book,RovSme} has proposed the `thermal time hypothesis' to address this question. The essential idea is a link between a local statistical state and a dynamical flow on physical quantities. This breaks the relationship between the space time metric and temporality and thus looks promising for a concept of time in quantum gravity. The thermodynamics of the machines we call clocks gives a concrete manifestation of this idea. 

As discussed in this paper, periodic clocks are driven, non equilibrium dissipative systems.  Non periodic clocks (thermal clocks) are systems driven from thermal equilibrium by measurement induced entropy reduction. In both cases they require a source of free energy to operate; work in the case of periodic clocks and accurate measurement in the case of non periodic clocks. All clocks are irreversible systems. 

In the case of thermal clocks the steady state of the unconditional (non measured) dynamics is given by the Gibbs-Boltzmann distribution
\begin{equation}
    \rho_\beta= Z^{-1} e^{-\beta \hat{H}}
\end{equation}
where $\beta=1/k_BT$ and $\hat{H}$ is the Hamiltonian operator.
( I will assume quantum  physics here). Proceeding formally for now, take the log of $\rho$ as a definition of the Hamiltonian up to a multiplicative factor
\begin{equation}
    \beta \hat{H} =\ln(\rho_\beta)
\end{equation}
The thermal flow of an observable represented by a hermitian operator  $\hat{A}$  defined by $\alpha^{\rho}_\tau(\hat{A})$   satisfies
\begin{equation}
    \frac{d\alpha_\tau^\rho(\hat{A})}{d\tau} = -\frac{i}{\hbar}[\log\rho,\hat{A} ]
\end{equation}
The thermal time of an equilibrium state $\tau$ at temperature $T$ and the Schroedinger quantum mechanical  time $t$ are related by 
\begin{equation}
\frac{d}{d\tau}= \beta \frac{d}{dt}
\end{equation}
Rovelli and Smerlak \cite{RovSme} interpret $\beta$ as the ``speed'' of thermal time; a rather novel reinterpretation of temperature. 

We can obtain a similar result by assuming that the history of thermal jumps $t_n$ in the thermal clock index the state of every other local system, In other words the state of a system is parameterised in terms of the stochastic process $\rho_S(t_n)$. One way to think of this is that we only ever measure the system state at the times $t_n$. In \cite{ID} I showed how this kind of stochastic time can be formulated as a dynamical flow in terms of  local lab time. If we average over the stochastic history of times, the dynamics can be written in Schroedinger mechanical time as 
\begin{equation}
\label{milburn-eqn}
    \frac{d\rho_s}{dt}= \gamma\bar{n}\left ( \exp\left [-i\frac{ \hat{H}_s}{\hbar\gamma\bar{n}}\right ]\rho_s\exp\left [i\frac{ \hat{H}_s}{\hbar\gamma\bar{n}}\right ]-\rho_s\right  )
\end{equation}
At high temperatures $\bar{n}=(\beta\epsilon)^{-1}$ where $\epsilon$ is the energy difference between the two states of the thermal clock. We then find that in the limit $\gamma \rightarrow \infty$, the good clock limit,  the dynamics may be approximated by 
\begin{equation}
\frac{d\rho_s}{d\tau}=\frac{-i}{\hbar}[\ln\rho_{s,\beta},\rho_s]
\end{equation}
where $\rho_{s,\beta}$ is the steady state of the system itself at thermal equilibrium with the same environment as that of the clock. This is analogous to Rovelli's thermal time flow differential map. It must be admitted however that the first order corrections to this equation in $\gamma^{-1}$ lead to intrinsic decoherence in all quantum dynamics\cite{ID}.  

Rovelli actually makes a more general hypothesis: any stationary state $\rho$ can be used to define a temporal flow. I expect that this corresponds to the more general case of irreversible periodic clocks.
It is interesting to note that the stationary solutions of Eq.(\ref{milburn-eqn}) are diagonal in the eigenstates of the system Hamiltonian. This would justify applying the thermal time hypothesis in the more general form.

\section{Conclusion.}
All physical clocks are open non equilibrium systems constrained by thermodynamics  and necessarily irreversible. All clocks are physical systems pushed away from thermal equilibrium thereby increasing their free energy. In the case of periodic clocks this is achieved by an external mechanism doing work on the system, for example a battery.  In the case of non periodic `thermal clocks' it is achieved by decreasing entropy through thermal or chemical potential gradients or, more interestingly, through measurement.  

As dissipative systems, all clocks are subject to noise which limits their performance. For classical clocks the source of noise is thermal and goes to zero with temperature. Thermal noise leads to phase diffusion when clocks are running on large amplitude limit cycles.   In the case  of quantum  clocks noise remains even at zero temperature due to, for example,  spontaneous emission or quantum tunnelling. By using a range  of examples both classical and quantum I have shown that the thermodynamic constraints require good clocks to dissipate large amounts of energy. 

An understanding of the thermodynamic limits of clocks gives important insights into the nature of physical time. I have shown how, for example, thermal clocks can be used to give a physical realisation of Rovelli's thermal time hypothesis. 

Our understanding of clock performance is crucial for emerging quantum technologies\cite{Ball}. Indeed ideas from quantum information, such as gate fidelity, may provide even better measures of clock performance than traditional metrics.  As our understanding of the links between thermodynamics and quantum information increases we  will be led to novel kinds of clocks that exceed the already incredible performance of current atomic clocks.

\section*{Acknowledgement}
This work was partly supported by a grant from FQXI for the program, ``Information as fuel", and the  Australian Research Council Centre of Excellence for Engineered Quantum Systems (Project number CE170100009). I would like to thank the following people for useful discussions: Thomas Milburn, Nicolas Meniccuci, Huw Price, Natalia Ares, Arkady Fedorov, Marcus Huber and Sally Shrapnel.

\end{document}